\documentclass[a4paper,11pt,prd,preprint, superscriptaddress, amsfonts,amsmath,amssymb,nofootinbib]{revtex4}
\usepackage[margin=20truemm]{geometry}
\usepackage{bm}
\usepackage[dvipdfmx]{graphicx}
\usepackage{ascmac}
\usepackage{amsthm}
\usepackage{latexsym}
\usepackage{physics}
\usepackage{color,float}
\usepackage{hyperref}
\usepackage{mathtools}
\usepackage{simplewick}
\usepackage{tikz}
\usepackage{adjustbox}
\usepackage{fancybox}
\usepackage{caption}
\usepackage{subcaption} 
\usepackage{tikz-feynman}
\usepackage{enumitem}
\usepackage{titlesec}
\allowdisplaybreaks

\begin{document}
\title{Gravitational effects on a dissipative two-level atom in the weak-field regime}

\author{Kaito Kashiwagi} 
\email{kashiwagi.kaito.268@s.kyushu-u.ac.jp}
\affiliation{Department of Physics, Kyushu University, Fukuoka, 819-0395, Japan}
\author{Akira Matsumura}
\email{matsumura.akira@phys.kyushu-u.ac.jp}
\affiliation{Department of Physics, Kyushu University, Fukuoka, 819-0395, Japan}
\affiliation{Quantum and Spacetime Research Institute, Kyushu University, 744 Motooka, Nishi-Ku, Fukuoka 819-0395, Japan}

\begin{abstract}
    We investigate
    the dissipative dynamics of a two-level atom in a weak gravitational field. 
    Using the Feynman--Vernon influence functional formalism, we derive a quantum master equation describing the two-level 
    atom interacting with a scalar field in a Newtonian gravitational field, and compute the energy dissipation rate of the atom.
    We find that the spontaneous emission rate (the dissipation rate in vacuum) is modified by the gravitational field. Specifically, this modification depends on the atom's dipole, the position of the atom relative to the source of the gravitational field, and the frequency of the scalar radiation emitted by the atom.
    Furthermore, we identify the parameter regimes in which the spontaneous emission rate is enhanced or suppressed by gravity.
    We also discuss how the modification arises from time dilation and dipole radiation in a weak gravitational field.
    These findings provide a theoretical basis for exploring gravitational effects in open quantum systems.
\end{abstract}

\maketitle
\tableofcontents
\section{Introduction}
\label{sec:Intro}
General relativity (GR)
and quantum mechanics (QM)
are two fundamental pillars of modern physics. 
GR describes the macroscopic scale physics and it is experimentally confirmed that this theory can explain various phenomena such as the perihelion precession of Mercury, gravitational lensing and gravitational redshift and so on\cite{Walsh_1979,Pound_1959}.
Especially, since the observation of gravitational waves(GRs) in 2015\cite{GWs_2016}, the observations of GWs have provided further evidences for the robustness of GR\cite{GR_test1,GR_test2}.
On the other hand, QM successfully describes microscopic dynamics, and abundant experimental evidence supports its validity, including the Stern--Gerlach experiment and the double-slit experiment \cite{SG_test, Double_slit_test}. 
More recently, the violation of Bell's inequality\cite{Bell_test1,Bell_test2,Bell_test3} and macroscopic quantum tunneling\cite{Q_tunnel_1985} have been experimentally observed.
Although it remains unclear whether gravity itself should be quantized, the effects of classical gravitational fields on quantum systems have been both theoretically investigated and experimentally verified. 
For example, as the well-known experiments, there are the measurement of the gravitationally induced phase shift of neutrons and the test of the quantized bound energy of neutron in the Earth's gravitational potential~\cite{COW_1975,Neutron_2002} and so on.

Rapid advances in quantum technologies, such as quantum sensing and quantum control, have made it possible to realize quantum systems and to detect extremely weak external influences.
For instance, quantum control over atomic ensembles containing $N \ge 10^8$ atoms has been demonstrated~\cite{QC_1,QC_2,QC_3}, and large-scale quantum entanglement among atoms has also been reported~\cite{QE_1,QE_2}.
In the field of quantum sensing, magnetic-field detection beyond the standard quantum limit has been achieved through quantum nondemolition measurements with laser-cooled rubidium atoms~\cite{QS_1}.
Furthermore, quantum-logic techniques applied to nitrogen-vacancy spin ensembles in diamond have enabled significant enhancement of magnetic-field sensitivity~\cite{QS_2}.
These developments open a new avenue toward exploring subtle gravitational effects using quantum systems.
In this study, we focus on the dissipative dynamics of a two-level atom 
and theoretically investigate how gravitational fields modify this dissipation process.
Since dissipation is closely tied to the exchange of information and energy between a quantum system and its environment, understanding how gravity influences such processes may offer new insights and possibilities for both fundamental physics and potential applications.

We consider that our present study leads to a model-independent method for Dark Matter (DM) detection.
The DM is thought to make up about 27\% of this universe, and the clarification of DM is one of the important issues in modern physics.
Various DM candidates such as Axion, Weakly Interacting Massive Particles(WIMPs) and Primordial Black Hole(PBH) have been considered, and those experimental verifications have been discussed \cite{DM_His_1,DM_His_2}.
For example, in the case of axions, the observation of the axion--photon conversion phenomenon, in which photons are converted into axions in the presence of a strong magnetic field, has been proposed \cite{DM_His_1,DM_Exp_1}. 
For WIMPs, experiments such as XENON1T aim to detect the recoil energy produced by their scattering with atomic nuclei \cite{DM_His_1,DM_Exp_2}.
Although these experiments have constrained the mass of DM and the coupling with DM, DM has yet to be directly detected. 
One possible reason of this situation is that the detection methods are based on specific model.
In light of the fact that DM was introduced as an unknown gravitational source~\cite{DM_His_1}, probing its gravitational interaction offers a model-independent approach. Nevertheless, because gravity is extremely weak compared with other fundamental interactions, highly sensitive detection is essential, and quantum technologies offer a promising route to achieve it.
For example, the detection of the gravitational impulse with heavy DM by quantum-mechanical sensors was proposed \cite{Dan_2020}.
Our study of the dissipative dynamics of atoms in a weak gravitational field potentially paves the way toward a model-independent and highly sensitive DM detection.

As another application, our study also contributes to high-precision tests of GR. 
Although GR has been extensively tested, probing possible deviations at small scales remains important.
Recent quantum technologies have enabled increasingly sensitive tests of GR using quantum systems.
For instance, Ref.~\cite{APJP_2025} demonstrated the high-precision measurement of gravitational acceleration through the gravity-induced dephasing of qubits,
and Ref.~\cite{EP_test} performed a precision test of the equivalence principle by observing gravitationally induced quantum interference with atom interferometry. 
Moreover, the GW detection using Rydberg atom and two-level atom were proposed\cite{GW_2-Lv1, GW_2-Lv2,GW_Rd_atom}. 
We consider that studying the dissipative dynamics of atoms in a gravitational field gives a route toward the tests of GR.

As mentioned above, our aim is to investigate the dissipative dynamics of a two-level atom as a pathway toward establishing a novel detection method in the interface between quantum physics and gravity. 
In this study, we focus on the energy dissipation of a two-level atom placed in a weak Newtonian gravitational field, in order to clarify how the gravitational field modifies the dissipative process. 
As a result, we find that the dissipation rate is suppressed or enhanced by the gravitational field, and that this behavior depends on the mass of gravitational source, the distance from the source to the atom, and the frequency of radiation field emitted by the atom. 

The structure of this paper is as follow.
In Sec.\ref{sec:Case_curve}, we derive the quantum master equation (QME) by using the Feynman-Vernon Influence functional method, which is one of methods that describes open quantum systems, and we analyze obtained the dissipation rate in the weak gravitational field in Sec.\ref{sec:analysis_Gamma}.
In Sec.\ref{sec:discuss}, we discuss the results of Sec.\ref{sec:Case_curve} and \ref{sec:analysis_Gamma}, and we finally conclude this study and mention the future outlook in Sec.\ref{sec:conclusion}.
In the following, we adopt the natural units $\hbar = c = 1$ and the convention of the Minkowski metric as 
$\eta_{\mu\nu} = \mathrm{diag}[-1, 1, 1, 1]$. 
The Newtonian gravitational constant is denoted by $G$.  
The commutator and the anticommutator are defined as 
$[\hat{A}, \hat{B}] = \hat{A}\hat{B} - \hat{B}\hat{A}$ 
and 
$\{\hat{A}, \hat{B}\} = \hat{A}\hat{B} + \hat{B}\hat{A}$, respectively. 
\section{QME of a two-level atom in a weak gravitational field}
\label{sec:Case_curve}
Quantum systems interacting with external environments are referred to as open quantum systems \cite{Breuer_2002}.
The dissipation of open quantum systems is one of the most fundamental and representative phenomena.
The dynamics can be systematically described within the theory of open quantum systems.
In this section, we derive the quantum master equation (QME) for a two-level atom in a weak gravitational field by employing the Feynman–Vernon influence functional method.
We begin with a brief overview of this method\cite{Hu_2008}.  
To this end, let us first write down the total action of the open quantum system and its environment:
\begin{align}
    S_\text{tot}
 &= S_\text{sys}[q] + S_\text{E}[\phi] + S_\text{int}[q,\phi] ,
    \label{eq:def_tot}
\end{align}
where $S_{\text{sys}}[q]$ denotes the action of the system of interest, 
the action of environment is described by $S_{\text{E}}[\phi]$, 
and $S_{\text{int}}[q,\phi]$ specifies the interaction between them.
QME is an equation of motion for the reduced density matrix $\rho_\text{s}(q, q',t)$ of the system, and this matrix can be yielded by using the Feynman-Vernon influence functional method.
Suppose that the density matrix in the total system is written as $\rho(q ,\phi , q', \phi',t)$, and this matrix element is given by the path integral,
\begin{align}
    &\rho(q ,\phi , q' ,\phi', t) \notag \\
    &= \int dq_{0} \int d\phi_{0}  \int dq_{0}' \int d\phi_{0}'
    \int^q_{q_{0}} \mathcal{D}q \int^\phi_{\phi_{0}} \mathcal{D}\phi
    \int^{q'}_{q_{0}'} \mathcal{D}q' \int^{\phi'}_{\phi_{0}'} \mathcal{D}\phi' \ e^{i(S_\text{tot}[q,\phi] - S_\text{tot}[q',\phi'])}
    \rho(q_{0},\phi_{0}, q_{0}',\phi_{0}',0). 
    \label{eq:def_rho_tot}
\end{align} 
The reduced density matrix $\rho_\text{s}$ is obtained by tracing over the environmental degrees of freedom $\phi$ in the total density matrix.
When the initial state $\rho(q_{0},\phi_{0}, q_{0}',\phi_{0}',0)$ is not correlated, that is, $\rho(q_{0},\phi_{0}, q_{0}',\phi_{0}',0) = \rho_\text{s}(q_{0},q_{0}',0)\rho_\text{E}(\phi_{0},\phi_{0}',0)$, we get the following form of $\rho_\text{s}$ as 
\begin{align}
    \rho_\text{s}(q,q',t) &= \int d\phi \ \rho(q,\phi,q',\phi,t) \notag \\
    &= \int dq_{0} \int dq_{0}' \ \rho_\text{s}(q_{0},q_{0}',0)
    \int^q_{q_{0}} \mathcal{D}q \int^{q'}_{q_{0}'} \mathcal{D}q' \ e^{i \left( S_\text{sys}[q] - S_\text{sys}[q'] + S_\text{IF}[q,q']\right) }, 
    \label{eq:re_rho1}
\end{align}
where the action $S_\text{IF}[q,q']$ is called the influence action defined by 
\begin{align}
    e^{iS_\text{IF}[q,q']} = \int d\phi \int d\phi_{0}  \ \int d\phi_{0}'
    \int^\phi_{\phi_{0}} \mathcal{D}\phi
    \int^{\phi}_{\phi_{0}'} \mathcal{D}\phi' \ e^{i(S_\text{E}+S_\text{int}[q, \phi]-S_\text{E}[\phi']-S_\text{int}[q',\phi'])} 
    \rho_\text{E}(\phi_{0}, \phi_{0}',0). 
    \label{eq:def_SIF0}
\end{align}
The influence action $S_\text{IF}[q,q']$ captures the effect from the environment. 
The explicit expression of $S_\text{IF}[q,q']$ makes it possible to describe the dynamics of the system of interest.

Next, we present the concrete setup in this paper.  
We consider that a gravitational source with a mass $M$ weakly bends the spacetime in the vicinity of the source. The spacetime metric is given by
\begin{align}
    g_{\mu\nu}(x) = 
    \text{diag}
    [-1 - 2\Phi(\bm{x}), 
    1 - 2\Phi(\bm{x}), 
    1 - 2\Phi(\bm{x}), 
    1 - 2\Phi(\bm{x})],
    \label{eq:def_metric}
\end{align}
where 
$\Phi(\bm{x})=-GM/x$ is the gravitational potential produced by the source, and $x=|\bm{x}|$ is the distance from the source. 
We also consider a composite system of two point-like particles bound together, which will later be treated as a two-level atom.
We assume that the particles are non-relativistic and that the center-of-mass (COM) motion of the composite system is almost at rest.
The internal motion of the system is nothing but the relative motion of the particles, whose action is given as follows \cite{Zych_2019}: 
\begin{align}
    S_\text{sys}[\bm{r}]
    &= \int dt 
    \left[ 
    \frac{1}{2} 
    \left[
    1 -3 \Phi(\bm{R})
    \right]
    \mu  \dot{\bm{r}}^2(t) 
    +  
    \left[
    1 + 2\Phi(\bm{R})
    \right]
    \frac{\alpha^2}{4\pi r(t)}  \right],
    \label{eq:S_sys2}
\end{align}
where $\mu=m_1 m_2/(m_1+m_2)$ is the reduced mass of the two particles, and $\bm{R}$ and $\bm{r}(t)$ denote the COM and relative coordinates, respectively. The $1/r$ binding potential between the two particles characterizes the internal energy level of the composite system, and $\alpha$ is the coupling constant. 
To get the action \eqref{eq:S_sys2}, the contribution of spacetime curvature was ignored by assuming $|\bm{r}| \ll |\bm{R}|$.
The two particles are taken to be coupled with a massless scalar quantum field $\phi(x)$, whose free action is
\begin{align}
    S_\text{E}[\phi] 
    &= - \frac{1}{2} \int d^4x \, \sqrt{-g(x)} 
    g^{\mu\nu}(x)\nabla_\mu \phi(x)\nabla_\nu \phi(x), 
    \label{eq:def_S_E_curve}
\end{align}  
where $g$ is the determinant of $g_{\mu\nu}$, $g^{\mu\nu}$ is the inverse of $g_{\mu\nu}$ and $\nabla_\mu$ is the covariant derivative in this spacetime. 
The interaction between the particles and the scalar field is introduced as a linear coupling of the field $\phi(x)$ to an external source $J(\bm{q};x)$ induced by the particle trajectories Refs.\cite{Iso_2011,Oshita_2016}:
\begin{align}
    S_\text{int}[\bm{q},\phi] 
    &= \int d^4x \, \sqrt{-g(x)} \, \phi(x) \, J(\bm{q};x) ,
    \label{eq:def_S_int_curve}
\end{align}
with the source defined by
\begin{align} 
    J(\bm{q};x) 
    &\equiv 
    \frac{1}{\sqrt{-g(x)}} \sum_{i=1}^{2}
    \lambda_i \ \frac{d\tau_i}{dt}  \, 
    \delta^3\!\big(\bm{x} - \bm{q}_i(t)\big) ,
    \label{eq:def_Jbar}
\end{align}
where $\lambda_i$ denotes the coupling constant, and $\tau_i$ is the proper time of each particle. 
Note that $\bm{q}_i$ denotes the position of each particle.
Since the gravitational field is weak, we can perturbatively evaluate the interaction action $S_\text{int}$ with respect to the gravitaional potential $\Phi$.  
The factor $dt/d\tau_i$ in Eq.\eqref{eq:def_Jbar} is expanded as $d\tau_i \approx [1 + \Phi(t,\bm{q}_i(t)] dt$, and the source $J(\bm{q};x)$ is evaluated as
\begin{align}
    \sqrt{-g(x)}J(\bm{q};x)
    &\approx  
    \sum^2_{i=1} \left[1 + \Phi(\bm{q}_i) \right]
    \lambda_i \delta^3( \bm{x} - \bm{q}_i(t) ). 
    \label{eq:def_Jbar2}
\end{align}
Substituting this into \eqref{eq:def_S_int_curve}, we have 
\begin{align}
    S_\text{int}
    &\approx \int dt
    \sum^2_{i=1} \left[1 + \Phi(\bm{q}_i) \right]
    \lambda_i \phi(t,\bm{q}_i). 
    \label{eq:S_int2}
\end{align}
Setting $\lambda = \lambda_1 = - \lambda_2$, we can realize a dipole-like interaction between the composite system and the scalar field. 
When the relative motion of the particles is small, using the relations $\bm{q}_1 = \bm{R}+ m_2 \bm{r} /(m_1 +m_2)$ and $\bm{q}_2 = \bm{R}- m_1 \bm{r}/(m_1 +m_2)$, we can expand the interaction as  
\begin{align}
    S_\text{int}
    &\approx [1+\Phi(\bm{R})]\int dt
    \, \lambda \bm{r}\cdot \nabla_x  \phi(t,\bm{R}),
    \label{eq:S_int3}
\end{align}
where we ignored the spatial derivative of the gravitational potential by assuming $|\bm{r}| \ll |\bm{R}|$. 
This is similar to the dipole interaction $-\bm{d} \cdot \bm{E}$ between a dipole and an electric field.  

Here, we regard the scalar field as the environment coupled with the composite system.
We also assume that the scalar field is in a thermal state in the weak gravitational field.
Tracing over the scalar field, based on Ref.\cite{Hu_2008}, we get the influence action for the composite system as
\begin{align}
    S_\text{IF}[\bm{r},\bm{r}']
    &=
    \frac{\lambda^2}{2} [1+\Phi(\bm{R})]^2\int dt \int dt'
    \left[
   r_i (t) - r'_i (t) 
    \right]
   \partial^x_i\partial^{x'}_j \mathbb{D}(t,\bm{R},t',\bm{R})
    \left[
    r_j(t') + r'_j (t')
    \right]
    \nonumber 
    \\
    &+ \frac{i\lambda^2}{2} [1+\Phi(\bm{R})]^2\int dt \int dt' \
    \left[ 
    r_i(t) - r'_i(t) 
    \right]
    \partial^x_i\partial^{x'}_j\mathbb{N}(t,\bm{R},t',\bm{R})
    \left[ 
   r_j (t') - r'_j (t') 
    \right],
    \label{eq:IF}
\end{align}
where $i$, $j$ running over spatial indices are implicitly summed over, and  
\begin{align}
    \mathbb{D}(t,\bm{x},t',\bm{x}') \equiv i  \langle[\hat{\phi}(t,\bm{x}), \hat{\phi}(t',\bm{x}')] \rangle \theta(t-t'), \quad \mathbb{N}(t,\bm{x},t',\bm{x}') \equiv \frac{1}{2} \langle \{\hat{\phi}(t,\bm{x}), \hat{\phi}(t',\bm{x}')\} \rangle.
    \label{eq:def_DN}
\end{align}
The bracket $\langle \cdot \rangle$ represents the expectation value taken for the thermal state of the scalar field. 
Explicitly, the functions $\mathbb{D}$ and $\mathbb{N}$ is perturbatively evaluated as
\begin{equation}
\mathbb{D} \approx \mathbb{D}_0 +\delta \mathbb{D}
, \quad \mathbb{N}\approx \mathbb{N}_0 +\delta \mathbb{N}, 
\label{eq:DN}
\end{equation}
where 
\begin{align}
    \mathbb{D}_0(t,\bm{x},t',\bm{x}')
    &=
    i\theta(t - t')
    \int \frac{d^3p}{(2\pi)^3} \frac{1}{2|\bm{p}|}
    e^{i \bm{p} \cdot (\bm{x} - \bm{x'})} I^{-}_p(t,t')
    \label{eq:def_D0}
    \\
    \delta \mathbb{D}(t,\bm{x},t',\bm{x}')
    &=
     8 \pi i R \Phi (\bm{R}) \theta(t - t')
    \int \frac{d^3p}{(2\pi)^3} \int \frac{d^3k}{(2\pi)^3}
    \frac{e^{ i \bm{p} \cdot \bm{x} - i \bm{k} \cdot \bm{x'} }}
    { |\bm{p} - \bm{k}|^2 ( |\bm{p}|^2 - |\bm{k}|^2) }
    \left[
    |\bm{p}| I^{-}_p(t,t') - |\bm{k}| I^{-}_k(t,t')
    \right]
    \label{eq:def_deltaD}
    \\
    \mathbb{N}_0 (t,\bm{x},t',\bm{x}')
    &=
    \frac{1}{2} 
    \int \frac{d^3p}{(2\pi)^3} \frac{1}{2|\bm{p}|}
    e^{i \bm{p} \cdot (\bm{x} - \bm{x'})} I^{+}_p(t,t')
    \label{eq:def_N0}
    \\
    \delta \mathbb{N} (t,\bm{x},t',\bm{x}')
    &=
    4 \pi R \Phi (\bm{R}) 
    \int \frac{d^3p}{(2\pi)^3} \int \frac{d^3k}{(2\pi)^3}
    \frac{e^{ i \bm{p} \cdot \bm{x} - i \bm{k} \cdot \bm{x'} }}
    { |\bm{p} - \bm{k}|^2 ( |\bm{p}|^2 - |\bm{k}|^2) }
    \left[
    |\bm{p}| I^{+}_p(t,t') - |\bm{k}| I^{+}_k(t,t')
    \right]
    \label{eq:def_deltaN}
    \\
    I^{-}_p(t,t')
    &\equiv
    e^{-i |\bm{p}| (t - t')}
    -
    e^{i |\bm{p}| (t - t')}
    ,
    \quad
    I^{+}_p(t,t')
    \equiv
    [ 2n_B( |\bm{p}| ) + 1 ]
    \left[
    e^{-i |\bm{p}| (t - t')}
    +
    e^{i |\bm{p}| (t - t')}
    \right]
    .
    \label{eq:def_I}
\end{align}
The $\mathbb{D}_0$ and $\mathbb{N}_0$
are computed from the thermal state of scalar field in the flat space, and the $\delta \mathbb{D}$ and $\delta \mathbb{N}$ correspond to those gravitational corrections. 
The mean occupation number of the thermal scalar particles is $n_B (|\bm{p}|)=1/(e^{\beta |\bm{p}|}-1)$ with the inverse temperature $\beta$.
The detailed derivation of Eqs.\eqref{eq:def_D0}, \eqref{eq:def_deltaD}, \eqref{eq:def_N0} and \eqref{eq:def_deltaN} is devoted in Appendix \ref{app:Analysys_two-point}.
Note that the functions, $\mathbb{D}_0$, $\mathbb{N}_0$, $\delta \mathbb{D}$ and $\delta \mathbb{N}$, remain valid when the wavelength of the scalar field is larger than the gravitational radius of the source, since they are obtained perturbatively by solving the wave equation of the field up to order $\mathcal{O}(\Phi)$.
From Eq.\eqref{eq:DN}, the influence functional up to the first order in the gravitational potential $\Phi$ is
\begin{align}
    S_\text{IF}[\bm{r},\bm{r}']
    &=
    \frac{\lambda^2}{2} \int dt \int dt'
    \left[
   r_i (t) - r'_i (t) 
    \right]
    \mathbb{D}_{ij}(t,t')
    \left[
    r_j(t') + r'_j (t')
    \right]
    \nonumber 
    \\
    &+ \frac{i\lambda^2}{2} \int dt \int dt' \
    \left[ 
    r_i(t) - r'_i(t) 
    \right]
    \mathbb{N}_{i j}(t, t')
    \left[ 
   r_j (t') - r'_j (t') 
    \right],
    \label{eq:IF2}
\end{align}
where    
\begin{align}
    \mathbb{D}_{ij}(t, t')
    &\equiv
   \left[ 
    1 + 2\Phi(\bm{R})
    \right]
    \partial^x_i \partial^{x'}_j 
    \mathbb{D}_0(t,\bm{R}, t',\bm{R}) 
    +
    \partial^x_i \partial^{x'}_j 
    \delta \mathbb{D}(t,\bm{R}, t', \bm{R}),
    \label{eq:def_D} 
    \\
    \mathbb{N}_{ij}(t,t')
    &\equiv
    \left[ 
    1 + 2\Phi(\bm{R})
    \right]
    \partial^{x}_i  \partial^{x'}_j
    \mathbb{N}_0(t, \bm{R}, t',\bm{R})
    +
    \partial^{x}_i  \partial^{x'}_j
    \delta \mathbb{N}(t,\bm{R},t', \bm{R}). 
    \label{eq:def_N}
\end{align}

The QME of the composite system is obtained as the time differential equation of the reduced density matrix of the system. 
From Eq.~\eqref{eq:re_rho1}, the reduced density matrix, 
$\rho_\text{s}(\bm{r},\bm{r}',t)$, is 
\begin{align}
    \rho_\text{s}(\bm{r},\bm{r}',t) 
    &= \int d\bm{r}_{0} \int d\bm{r}_{0}'\, \rho_\text{s}(\bm{r}_{0},\bm{r}_{0}',0) 
    \int_{\bm{r}_{0}}^{\bm{r}} \mathcal{D}\bm{r} 
    \int_{\bm{r}_{0}'}^{\bm{r}'} \mathcal{D}\bm{r}'\,
    e^{i( S_\text{sys}[\bm{r}] - S_\text{sys}[\bm{r}'] +S_\text{IF}[\bm{r},\bm{r}'])}.
    \label{eq:re_rho2}
\end{align}
Computing the time derivative of $\rho_\text{s}(\bm{r}, \bm{r}',t)$ and expanding it up to $O(\lambda^2)$ (employing the Born approximation), we derive the following QME for $\hat{\rho}_\text{s}(t)=\int d\bm{r} \int d\bm{r}' \rho_\text{s}(\bm{r}, \bm{r}',t)|\bm{r} \rangle \langle \bm{r}'|$,
\begin{align}
    \frac{\partial}{\partial t} \hat{\rho}_\text{s}(t)
    &=
    -i  
    \left[
    \hat{H}_s, \hat{\rho}_\text{s}(t)
    \right]
    +
    \frac{i \lambda^2}{2} \int^t_0 dt' \ 
    \mathbb{D}_{ij}(t,t') 
    \left[ 
    \hat{r}_i, 
    \left\{ 
    \hat{r}_j (t' - t), \hat{\rho}_\text{s}(t) 
    \right\}
    \right]
    \notag
    \\
    &\quad - \lambda^2 \int^t_0 dt' \ 
    \mathbb{N}_{ij}(t,t') 
    \left[ 
    \hat{r}_i, 
    \left[
    \hat{r}_j (t' - t), \hat{\rho}_\text{s}(t) 
    \right]
    \right],
    \label{eq:QME_2particle}
\end{align}
where $\hat{r}_j (t)=e^{i\hat{H}_\text{s} t} \hat{r}_j e^{-i\hat{H}_\text{s} t}$ is the operator of the relative coordinate, and the system Hamiltonian $\hat{H}_s$ is given by
\begin{align}
    \hat{H}_s
    =
    \left[
    1 + 3\Phi(\bm{R})
    \right]
    \ \frac{ \hat{\bm{p}}^2 }{ 2\mu }
    -
    \left[
    1 + 2\Phi(\bm{R})
    \right]
    \ \frac{\alpha^2}{4\pi \hat{r}} . 
    \label{eq:Hs_2particle}
\end{align}
The operator $\hat{\bm{p}}$ is the canonical momentum conjugate to the relative coordinate. 
For the detailed derivation of the QME, see Appendix \ref{app:Derivation_QME}. 

In the following, approximating the composite system as a two-level atom, we derive the QME for the two-level system. 
To identify the physical energy of the composite system, we express the Hamiltonian $\hat{H}_\text{s}$ in terms of proper quantities\cite{Zych_2019}.
Let $dt_{\text{p}}$ and $d\bm{x}_\text{p}$ denote the infinitesimal proper temporal and spatial intervals, respectively, defined in the local rest frame of the COM motion of the composite system. 
They are related to the coordinate intervals as 
$dt_{\text{p}} \approx [1 + \Phi(\bm{R})]\,dt$ and $d\bm{x}_\text{p}\approx [1 - \Phi(\bm{R})]\,d\bm{x}$.  
The conjugate momentum $\bm{p}$ computed from Eq.\eqref{eq:S_sys2} is rewritten as 
\begin{align}
    \bm{p} 
    &= 
    [1 - 3\Phi(\bm{R})]\mu \frac{d\bm{r}}{dt}
    \approx 
    [1 - \Phi(\bm{R})]\mu \frac{d\bm{r}_\text{p}}{dt_\text{p}}
    =
    [1 - \Phi (\bm{R})] \bm{p}_{\text{p}},
\end{align}
where $d\bm{r}_\text{p} \approx [1 - \Phi(\bm{R})]\,d\bm{r}$ and $\bm{p}_{\text{p}}=\mu \, d\bm{r}_\text{p}/dt_\text{p}$.
Given this result, the Hamiltonian $H_s$ is expressed as
\begin{align}
    H_s 
    &=
    [1 + 3\Phi(\bm{R})] \ \frac{\bm{p}^2}{2\mu}
    -
    [1 + 2\Phi(\bm{R})] \ \frac{\alpha^2}{4\pi r} 
    \notag \\
    &\approx
    [1 + 3\Phi(\bm{R})] \cdot [1 - 2\Phi(\bm{R})] \cdot \frac{\bm{p}^2_{\text{p}}}{2\mu}
    -
    [1 + 2\Phi(\bm{R})] \cdot [1 - \Phi(\bm{R})] \cdot \frac{\alpha^2}{4\pi r_{\text{p}} }
    \notag \\
    &\approx
    [1 + \Phi(\bm{R})] 
    \left[ \frac{ \bm{p}_{\text{p}}^2}{2\mu} 
    -  
    \frac{\alpha^2}{4\pi r_{\text{p}} }
    \right].
\end{align}
We assume that the internal state of the composite system is be in the energy eigenstates of the Hamiltonian $\hat{\bm{p}}_{\text{p}}^2/(2\mu) -\alpha^2/(4\pi \hat{r}_\text{p})$. 
Specifically, we pick up the ground state  
$|g\rangle$ and the excited state $|e\rangle$ with an energy splitting $\Omega_{\text{p}}$ measured in proper time. 
The Hamiltonian $\hat{H}_\text{s}$ can be approximated as
\begin{align}
    \hat{H}_\text{s} 
    \approx 
    \hat{H}^{(2)}_\text{s} 
    = 
    \frac{\Omega_g}{2} \hat{\sigma}_z 
    \quad 
    \Omega_g \equiv [1 + \Phi(\bm{R})] \Omega_{\text{p}},
    \label{eq:def_H_two-level}
\end{align}
where the Pauli Z is $\hat{\sigma}_z=|e\rangle \langle e|-|g\rangle\langle g|$.
Following the analogy of the dipole coupling with  electromagnetic fields, we also approximate $\lambda \hat{\bm{r}}_\text{p}$ as $\lambda \hat{\bm{r}}_\text{p} \approx \bm{d}_{\text{p}}  \hat{\sigma}_x$, where $\bm{d}_{\text{p}}$ denotes an effective proper dipole coupled to the scalar field, and  $\hat{\sigma}_x=|e\rangle \langle g|+|e\rangle\langle g|$. Then, $\lambda \hat{\bm{r}}$ in the terms with $\tilde{\mathbb{D}}_{ij}$ and $\tilde{\mathbb{N}}_{ij}$ of Eq.  \eqref{eq:QME_2particle} given as  
\begin{align}
    \lambda \hat{\bm{r}} \approx [1+\Phi(\bm{R})] \lambda \hat{\bm{r}}_\text{p} =[1+\Phi(\bm{R})]\bm{d}_{\text{p}}  \hat{\sigma}_x.
    \label{eq:lambda_r}
\end{align}
Hereafter, for simplicity, we omit the subscript p of $\Omega_{\text{p}}$ and $\bm{d}_{\text{p}}$, respectively.
Using \eqref{eq:def_H_two-level} and \eqref{eq:lambda_r}, the QME becomes
\begin{align}
    \frac{\partial}{\partial t} \hat{\rho}_\text{s}(t)
    &=
    -i  
    \left[
    \hat{H}^{(2)}_s, \hat{\rho}_\text{s}(t)
    \right] 
    +
    \frac{i}{2}  \int^t_0 dt' \ 
    d_i d_j \ \tilde{\mathbb{D}}_{ij}(t,t') 
    \left[ 
    \hat{\sigma}_x, 
    \left\{ 
    \hat{\sigma}_x(t' - t), \hat{\rho}_\text{s}(t) 
    \right\}
    \right]
    \notag \\
    &\quad
    - 
    \int^t_0 dt' \ 
    d_i d_j \ \tilde{\mathbb{N}}_{ij}(t,t') 
    \left[ 
    \hat{\sigma}_x, 
    \left[
    \hat{\sigma}_x(t' - t), \hat{\rho}_\text{s}(t) 
    \right]
    \right],
     \label{eq:QME_two-level}
\end{align}
where
 $\tilde{\mathbb{D}}_{ij}$ and $\tilde{\mathbb{N}}_{ij}$ were defined as
\begin{align}
    \tilde{\mathbb{D}}_{ij}(t,t')
    &\equiv
    [ 1 + 4\Phi(\bm{R}) ]
    \partial^x_i  \partial^{x'}_j
    \mathbb{D}_0(t,\bm{R}, t',\bm{R})
    +
    \partial^x_i  \partial^{x'}_j
    \delta \mathbb{D}(t,\bm{R}, t',\bm
    {R}), 
    \label{eq:tilD}
    \\
    \tilde{\mathbb{N}}_{ij}(t,t')
    &\equiv
    [ 1 + 4\Phi(\bm{R}) ]
    \partial^x_i  \partial^{x'}_j
    \mathbb{N}_0(t,\bm{R}, t',\bm{R})
    +
    \partial^x_i  \partial^{x'}_j
    \delta \mathbb{N}(t,\bm{R},t',\bm{R})
    \label{eq:tilN}.
\end{align}
We can cast Eq.\eqref{eq:QME_two-level} into the Gorini–Kossakowski–Sudarshan–Lindblad (GKSL) form by the Markovian and rotating-wave approximations\cite{Breuer_2002}. 
Practically, the Markovian approximation is to perform the variable transformation $t' - t \equiv - \tau$ and then to change the integral upper $t$ to $\infty$. 
The rotating-wave approximation is to ignore the terms proportional to $e^{\pm i\Omega_g t}$. 
Working in the interaction picture defined by the Hamiltonian $\hat{H}_\text{s}$ and applying those approximations, we arrive at the Markovian QME for the two-level atom
\begin{align}
  \frac{\partial}{\partial t} \hat{\rho}^\text{I}_\text{s}(t)
  = -i \left[ \hat{H}_\text{shift}, \hat{\rho}^\text{I}_\text{s}(t) \right] 
  + \, \mathcal{D}_{-}\left[ \hat{\rho}^\text{I}_\text{s}(t) \right] 
  + \mathcal{D}_{+}\left[ \hat{\rho}^\text{I}_\text{s}(t) \right],
  \label{eq:QME_Int2}
\end{align}
with $\hat{H}_\text{shift} 
  \equiv 
  S_- \hat{\sigma}_{+} \hat{\sigma}_- - S_+ \hat{\sigma}_{-} \hat{\sigma}_+$ and the dissipators
\begin{align}
  \mathcal{D}_{-}\left[ \hat{\rho} \right] \equiv \Gamma_{-} \Big[\hat{\sigma}_- \hat{\rho} \hat{\sigma}_+ - \frac{1}{2} \left\{ \hat{\sigma}_+ \hat{\sigma}_-, \hat{\rho} \right\}\Big], \quad 
  \mathcal{D}_{+}\left[ \hat{\rho} \right] \equiv \Gamma_+ \Big[\hat{\sigma}_+ \hat{\rho} \hat{\sigma}_- - \frac{1}{2} \left\{ \hat{\sigma}_- \hat{\sigma}_+,\hat{\rho} \right\} \Big],
  \label{eq:D_em_and_ab}
\end{align}
where 
\begin{align}
   &\Gamma_{\pm} =2\int^{\infty}_0 d\tau d_i d_j \, \text{Re} \Big[\Big( \tilde{\mathbb{N}}_{ij}(t,t-\tau)\pm
    \frac{i}{2} \tilde{\mathbb{D}}_{ij}(t,t-\tau) \Big) e^{i\Omega_g \tau} \Big] ,
    \label{eq:Gamma_pm}
    \\
  &S_{\pm} =\int^{\infty}_0 d\tau d_i d_j \, \text{Im} \Big[\Big( \tilde{\mathbb{N}}_{ij}(t,t-\tau) \pm
    \frac{i}{2} \tilde{\mathbb{D}}_{ij}(t,t-\tau) \Big) e^{i\Omega_g \tau} \Big], 
    \label{eq:S_pm}
\end{align} 
and the superscript $\text{I}$ of $\hat{\rho}^\text{I}_\text{s}$ indicates that Eq.\eqref{eq:QME_Int2} is written in the interaction picture, and $\hat{\sigma}_\pm \equiv (\hat{\sigma}_x \pm i \hat{\sigma}_y)/2$ are the ladder operators.

As shown in Eq.~\eqref{eq:QME_Int2}, the first term of Eq.~\eqref{eq:QME_Int2} with $\hat{H}_{\text{shift}}$ represents an environment-induced energy shift of the system.  This effect is known as the Lamb shift. Although the formal expression for $\hat{H}_{\text{shift}}$ is divergent, the divergence can be removed by renormalization.  
In this work, we treat this shift as a subleading contribution and therefore neglect it.  
Here we emphasize that the result in Eq.~\eqref{eq:QME_Int2} cannot be obtained by a simple gravitational redshift replacement $\Omega \rightarrow \Omega_g$, indicating that the environmental effects play an essential and nontrivial role.

\section{Dissipation rate in a weak gravitational field}
\label{sec:analysis_Gamma}

Since we have the explicit QME of the two-level atom, we can obtain the energy dissipation rate of the atom.
In order to derive the energy dissipation rate, we solve the QME~\eqref{eq:QME_Int2} (see Appendix \ref{app:solution_QME}).
For the initial condition, $\langle e|\hat{\rho}^\text{I}_\text{s}(0)|e\rangle=1$, $\langle e|\hat{\rho}^\text{I}_\text{s}(0)|g\rangle=0= \langle g|\hat{\rho}^\text{I}_\text{s}(0)|e\rangle=\langle g|\hat{\rho}^\text{I}_\text{s}(0)|g\rangle$, that is, 
\begin{equation}
    \hat{\rho}^\text{I}_\text{s}(0) = 
    \begin{pmatrix}
        1     &  0 \\
        0 &  0
    \end{pmatrix},
    \label{eq:rho_initial}
\end{equation}
the solution of the QME~\eqref{eq:QME_Int2} can be written as
\begin{align}
    \hat{\rho}^\text{I}_\text{s}(t) &=
    \begin{pmatrix}
        (1- a_s) e^{- \Gamma  t} + a_s & 0 \\
        0 & (1 - a_s)(1-e^{- \Gamma t})
    \end{pmatrix},
    \label{eq:solve_QME}
\end{align}
where 
\begin{align}
    \Gamma
    \equiv
    \Gamma_+ + \Gamma_-,
    \quad 
    a_s \equiv
    \frac{\Gamma_+ }{\Gamma_+ +\Gamma_-}, 
    \label{eq:def_gammaas}
\end{align}
and $\hat{H}_\text{shift}$ has been neglected as mentioned above.  
From Eq.~\eqref{eq:solve_QME}, we read out that the characteristic timescale of the dissipation is 
$1 / \Gamma$.  
In this section, we analyze $\Gamma$ in detail.

\begin{figure}[H]
    \centering
    \includegraphics[width=0.8\linewidth]{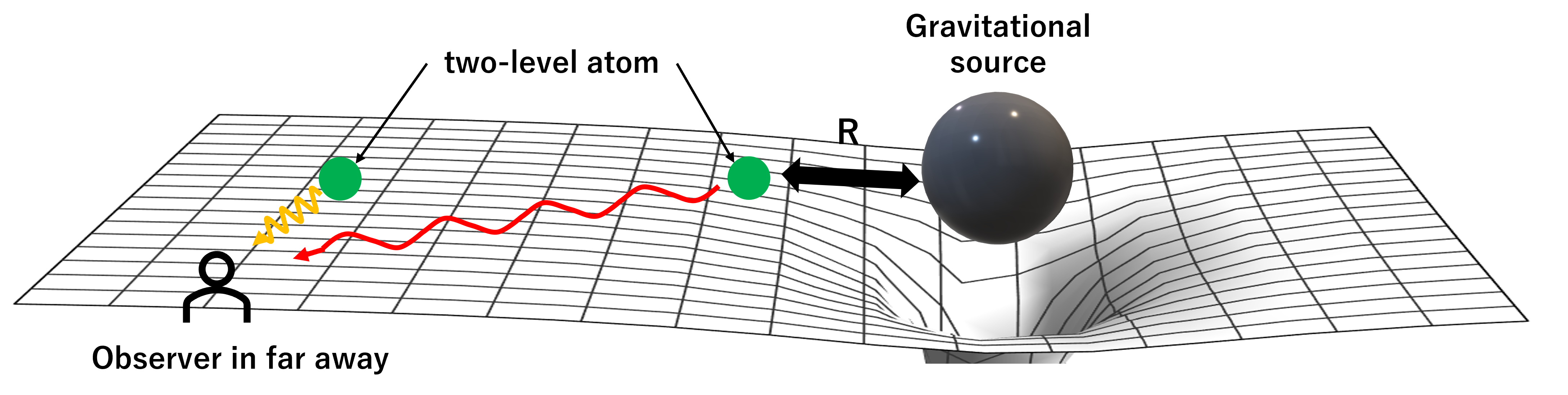}
    \caption{
    Schematic thought experiment illustrating how a distant  observer compares decay signals from two atoms: one  located near a gravitational source and another located  near the distant observer.
    By repeatedly detecting decay events from the distant  atom, the observer can infer its dissipation rate and compare it with the reference rate obtained from the atom, thereby identifying gravitational effects.
    }
    \label{fig:illust_situation}
\end{figure}
\noindent
Before the analysis, we discuss how the dissipation rate and its gravitational correction can be observed. 
To this end, let us consider the thought experiment illustrated in Fig.~\ref{fig:illust_situation}.
A distant observer far from the gravitational source measures physical quantities and can examine a two-level atom located near the observer. 
In this case, the dissipation rate directly measured by the observer coincides with the rate defined in the atom’s proper frame.
Next, we consider another identical atom placed near the gravitational source. 
The distant observer also can detect the scalar particles emitted by that atom and infer  the dissipation rate from the observed emission rate. 
The observer can then compare the two rates: one associated with the atom near the observer and the other with the atom near the gravitational source.
From the comparison, the observer can identify the gravitational correction of the dissipation rate.

\begin{figure}[H]
    \centering
    \includegraphics[width=0.4\linewidth]{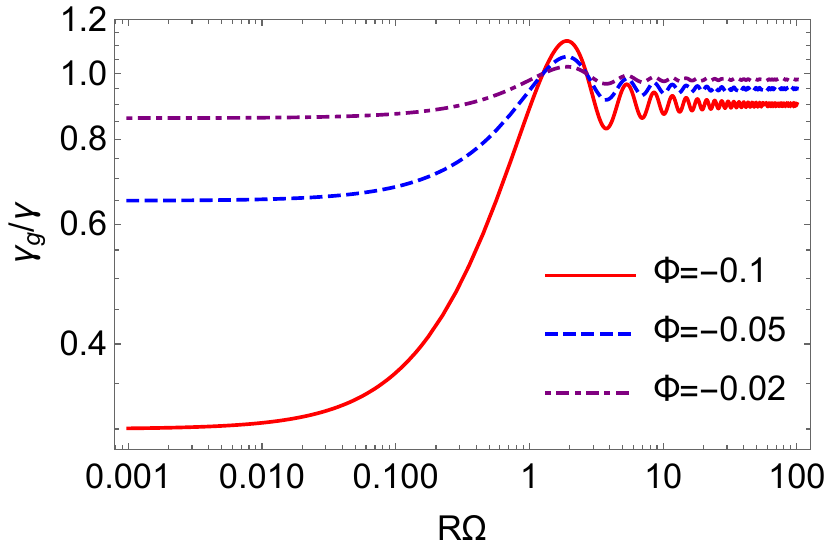}
    \hspace{1cm}
    \includegraphics[width=0.4\linewidth]{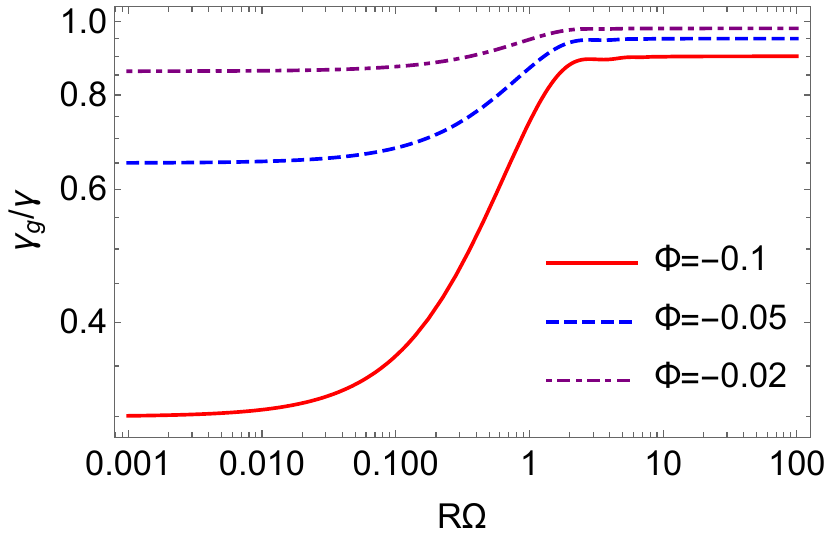}
    \caption{Behavior of the ratio $\gamma_g/\gamma$ as a function of $R\Omega$. The left panel is the parallel case $\bm{d}\parallel\bm{R}$, and the right panel is the perpendicular case $\bm{d}\perp\bm{R}$. 
    In the low- and high-frequency limits, the ratio becomes constant. 
    In the parallel case (the left panel), the ratio is greater than one around $R\Omega \sim 1$. 
    }
    \label{fig:ratio}
\end{figure}

To get the dissipation rate 
$\Gamma$, we calculate $\Gamma_+$ and $\Gamma_-$ given in Eq.\eqref{eq:Gamma_pm}. 
These are explicitly yielded as 
\begin{align}
  \Gamma_{+} =n_B(\Omega_g)\gamma_g, \quad 
    \Gamma_{-}=[n_B(\Omega_g)+1] \gamma_g, 
   \label{eq:Gamma+-}
\end{align}
where $n_B (\Omega_g)$ is the occupation number of the scalar particles at the frequency $\Omega_g$. The spontaneous emission rate $\gamma_g$ is 
\begin{align}
  \gamma_g
  = \gamma
    \left[
    1 + 7\Phi(\bm{R})
    -
    2\Phi(\bm{R})
     \, f_1(R\Omega)
    -
    3 \Phi(\bm{R})  \frac{(\bm{d} \cdot \bm{R} )^2 - d^2 R^2}{d^2 R^2 } f_2(R\Omega)
    \right],
    \label{eq:gamma_g}
\end{align}
where $\gamma=d^2\Omega^3/(6\pi)$, $R=|\bm{R}|$, $d=|\bm{d}|$, and the numeric functions, 
\begin{align}
 f_1(x) &\equiv
    \frac{1}{x^2} \Big[1 +x^2 ( \pi x + 3 ) 
    - ( 1 + x^2 ) \cos(2x)
    - 2 x \sin( 2 x )
    - 2 x^3 S_i(2 x)\Big],
    \label{eq:def_f1}
    \\
    f_2(x) 
    &\equiv
    \frac{1}{x^2}\Big[1 -x \sin(2 x) - \cos(2 x)\Big].
    \label{eq:def_f2}
\end{align}
The function $S_i(x)=\int^x_0 dy \sin y/y$ is the sine integral. 
The above $\Gamma_+$, $\Gamma_-$, and $\gamma_g$ are derived in Appendix \ref{app:detail_Ag-K}.
Here, $\gamma$ denotes the spontaneous emission rate of the two-level atom far from the gravitational source. 
Since gravitational effects are defined through the comparison of the two atoms illustrated in Fig\ref{fig:illust_situation}, the ratio $\gamma_g/\gamma$ provides a meaningful measure of the modification of the emission rate. 
Fig.\ref{fig:ratio} presents the behavior of $\gamma_g/\gamma$ as a function of $R\Omega$ for some fixed values of $\Phi$. 
The left and right panels show the ratio $\gamma_g/\gamma$ for the parallel case $\bm{d}\parallel\bm{R}$ and for the perpendicular case $\bm{d}\perp\bm{R}$, respectively. In the low- and high-frequency limits, the ratio gets to be constant. In the parallel case, the ratio is greater than one around $R\Omega \sim 1$. Hence the spontaneous emission rate $\gamma_g$ is enhanced. In the other parameter region, the rate is suppressed. 

Let us focus on the two limits: the low frequency limit  $R \Omega \ll 1$ (the long wavelength mode of emitted scalar particles) and the high frequency limit $R\Omega \gg 1$ (the short wavelength mode of emitted scalar particles).
Because the functions $f_1(x)$ and $f_2(x)$ are approximated as
\begin{equation}
   f_1(x) \approx
    \begin{cases}
        \pi x 
        &  (x \ll 1) \\[2pt]
        3
        & (x \gg 1)
    \end{cases}, \quad 
    f_2(x) \approx
    \begin{cases}
        \frac{4}{3} x^2
        &  (x \ll 1) \\[2pt]
        x \sin(2x)
        & (x \gg 1)
    \end{cases},
    \label{eq:f_approx}
\end{equation}
we obtain the following spontaneous emission rates in each regime:
\begin{equation}
    \gamma_g \approx 
    \begin{cases}
        \gamma [1 + 7\Phi(\bm R)], & (R\Omega\ll1),\\[4pt]
        \gamma [1 + \Phi(\bm R)], & (R\Omega\gg1). 
    \end{cases}
    \label{eq:summurize_gamma}
\end{equation}
Thus we find that the rate $\gamma_g$ is gravitationally reduced and that the reduction is different in each limit.
In the next section, we will discuss this behavior by focusing on gravitational time dilation and on dipole radiation in a weak gravitational field. 

\section{Discussion}
\label{sec:discuss}

Here, we discuss the peculiar nature of the spontaneous emission rate given in \eqref{eq:summurize_gamma}.
For this purpose, let us consider a distant observer who measures the dissipation of a two-level atom during a finite coordinate-time interval $\Delta t$.
In this setup, the quantity $\gamma_g \Delta t$ represents the number of the scalar particles emitted within the interval $\Delta t$. 
Eq.~\eqref{eq:summurize_gamma} may be written as
\begin{equation}
    \gamma_g  \Delta t  \approx
    \begin{cases}
        \left[ 1 + 6\Phi(\bm R) \right] \gamma  \Delta t_\text{p} 
        &  R\Omega \ll 1, \\[6pt]
         \, \gamma \Delta t_\text{p} \,
        &  R\Omega \gg 1,
    \end{cases}
    \label{eq:sum_gamma_eff}
\end{equation}
where $\Delta t_{\text{p}}$ is the proper temporal interval for the two-level atom and $\gamma$ is the proper spontaneous emission rate, respectively.
Hence, in the high-frequency limit, the spontaneous emission rate $\gamma_g$ in Eq.\eqref{eq:summurize_gamma} is explained by the invariance of the number of emitted particles $\gamma_g \Delta t=\gamma \Delta t_\text{p}$, together with gravitational time dilation $\Delta t_\mathrm{p} \approx (1 + \Phi)\Delta t$.
From the viewpoint of the equivalence principle, this behavior is consistent with GR, since the result can be understood in terms of gravitational redshift. 
In contrast, although the low-frequency result in Eq.~\eqref{eq:sum_gamma_eff} deviates from such an intuitive picture at first glance, this result does not means the violation of equivalence principle.
Because the Green function used in this paper is obtained perturbatively up to $\mathcal{O}(\Phi)$ and is valid for a wavelength larger than the gravitational radius of the source, this deviation indicates that the emission process is not determined solely by local physics, but is influenced by the nonlocal structure of field propagation and its gravitational modification encoded in the Green's function.
In addition, we observe that the result in Eq.~\eqref{eq:gamma_g} can also be understood from the viewpoint of energy balance. 
Since the energy of the two-level atom dissipates via scalar radiation emitted from the atom, one expects that the energy dissipation rate characterized by $\gamma_g$ can be estimated as $P/\Omega_g$, where $P$ is the power of the scalar radiation and $\Omega_g$ is the energy of the atom as measured by a distant observer. 
Indeed, as shown in Appendix~\ref{app:Derive_E-change}, this expectation is confirmed by analyzing the scalar radiation emitted from the atom's effective dipole, yielding
\begin{align}
    \frac{P}{\Omega_g} \approx \frac{1}{4} \gamma_g.
    \label{eq:E_change}
\end{align}
Furthermore, previous studies have reported that the energy carried by electromagnetic radiation emitted from charged particles receives contributions beyond a simple gravitational redshift~\cite{Peters_1973}, which is consistent with our results. 
We consider that the behavior of $\gamma_g$ in long-wavelength limit reflects a non-local effect of scalar field in a weak gravitational field.
The spontaneous emission rate $\gamma_g$ is  the dissipation rate in the vacuum environment.
Since spontaneous emission originates from vacuum fluctuations, the behavior in the long-wavelength regime is expected to be sensitive to how these fluctuations are modified by gravity and suggest that gravitational corrections to vacuum fluctuations may play an important role.
However, this point remains speculative and warrants further investigation in future work.

It is also important to discuss whether finite-temperature effects give the effects to the dissipation rate $\Gamma=\Gamma_+ + \Gamma_-$ in Eq.\eqref{eq:def_gammaas}.
In a finite-temperature environment, since we have 
$\Gamma_+ =n_B (\Omega_g) \gamma_g$ and $\Gamma_-=[n_B (\Omega_g)+1]\gamma_g$ according to Eq.\eqref{eq:Gamma+-}, 
the dissipation rate is 
\begin{align}
    \Gamma
    =[2 n_B (\Omega_g)+1] \gamma_g.
    \label{eq:Gamma_th}
\end{align}
Therefore, the thermal effect only comes from the Bose distribution $n_B(\Omega_g)=[ \exp[ \Omega_g/(k_B T)] - 1 ]^{-1}$.
The parameter $1/(k_B T)$, where $k_B$ is the Boltzmann constant, represents the inverse temperature assigned by a distant observer to the scalar field located at $\bm{R}$.
Here, we assume that an observer at the spatial infinity and another observer at $\bm{R}$ each prepare a finite-temperature scalar-field environment at the same “local" temperature ($= T_\text{p}$).
We then ask how the temperature at $\mathbf{R}$ is perceived by the observer at infinity, that is, ask the relation between $T$ and $T_\text{p}$.
In general relativity, the temperature of a system in thermal equilibrium depends on the background spacetime metric.  
In particular, in a stationary spacetime, the following relation called Tolman–Ehrenfest relation holds\cite{Tolman}:
\begin{align}
    T_\text{p}
    = \frac{T}{\sqrt{-g_{00}(\bm{R})}}
    \approx
    \frac{T}{1 + \Phi(\bm{R})}
    \label{eq:TE_relation}
\end{align}
Then, the Bose distribution $n_B(\Omega_g)$ can be rewritten as
\begin{align}
    n_B(\Omega_g) 
    = 
    \frac{1}{ \exp\left( \frac{\Omega_g}{k_B T} \right) - 1 }
    =
    \frac{1}{ \exp\left( \frac{\Omega}{k_B T_\text{p}} \right) - 1 }
    =
    n^\text{p}_B(\Omega ),
    \notag
\end{align}
so this result implies the distribution itself does not seem to have changed.

\section{Conclusion and outlook}
\label{sec:conclusion}

In this study, we investigated the energy-dissipative dynamics of a two-level atom in a weak gravitational field.
Using the Feynman--Vernon influence functional formalism, we derived a GKSL-form quantum master equation for the atom and evaluated its dissipation rate by solving the equation.
As a result, we found that gravitational modifications appear in the spontaneous emission rate, as shown in Eq.~\eqref{eq:gamma_g}.
Furthermore, by analyzing the behavior of the dissipation rate, we identified a parameter region in which the rate is enhanced, as illustrated in Fig.~\ref{fig:ratio}.
We also found that the reduction behavior of the rate differs between the low- and high-frequency limits, as summarized in Eq.~\eqref{eq:summurize_gamma}.
These results indicate that environmental effects play important and non-trivial roles.
In particular, we suggest that vacuum fluctuations of a scalar field in a weak gravitational field are key to understanding our results.

As mentioned in Sec.~\ref{sec:Intro}, several proposals have been made to detect gravitational effects using quantum systems.
With the remarkable progress of quantum sensing technologies in recent years, it is becoming increasingly feasible to detect even extremely weak gravitational influences.
If such experimental capabilities are realized, our research may provide a new theoretical pathway for the detection of gravitational waves and dark matter.
In addition, our results may also be applicable to tests of quantum field theory in curved spacetime, since it is considered that the phenomena discussed in this work are closely related to the properties of scalar fields, particularly their vacuum fluctuations.
We hope that our study may serve as a theoretical bridge toward the interface between quantum theory and gravity.

\begin{acknowledgements}
We thank Y. Chen and K. Yamamoto for variable discussions and comments related to this paper.
In addition, we thank the Yukawa Institute for Theoretical Physics at Kyoto University. Discussions during the YITP workshop YITP-W-25-11 were useful to complete this work.
K.K. was supported by JST SPRING (Japan Grant Number JPMJSP2136).
A.M. was supported by JSPS KAKENHI (Grants No.~JP23K13103 and No.~JP23H01175).
\end{acknowledgements}

\begin{appendix}

\section{Derivation of Eq.\eqref{eq:QME_2particle}}
\label{app:Derivation_QME}
In this appendix, we derive the QME of the composite system given in Eq.\eqref{eq:QME_2particle}.
We consider the infinitesimal time evolution from $t$ to $t+\epsilon$. At the time $t+\epsilon$, the reduced density matrix of the system, given by Eq.\eqref{eq:re_rho2}, is 
\begin{align}
    \rho_\text{s}(\bm{r},\bm{r}', t + \epsilon)
    &=
    \int d\bm{r}_{0} \int d\bm{r}_{0}'\,\rho_\text{s}(\bm{r}_{0},\bm{r}_{0}',0) 
    \int_{\bm{r}_0}^{\bm{r}} \mathcal{D}\bm{r} 
     \int_{\bm{r}'_0}^{\bm{r}'} \mathcal{D}\bm{r}'\,
    e^{i\left(S_\text{sys}[\bm{r},t + \epsilon] - S_\text{sys}[\bm{r}',t + \epsilon]\right)} e^{iS_\text{IF}[\bm{r},\bm{r}',t + \epsilon] } 
    \notag \\
    &\approx
    \int d\bm{r}_{0} \int d\bm{r}_{0}'\,\rho_\text{s}(\bm{r}_{0},\bm{r}_{0}',0) 
    \int_{\bm{r}_0}^{\bm{r}} \mathcal{D}\bm{r} 
     \int_{\bm{r}'_0}^{\bm{r}'} \mathcal{D}\bm{r}'
    \notag \\
    & \quad \times
    \left\{
    e^{i\left[S_\text{sys}[\bm{r},t + \epsilon] - S_\text{sys}[\bm{r}',t + \epsilon] \right]} 
    e^{iS_\text{IF}[\bm{r},\bm{r}',t]}
    +
    i \delta_\epsilon S_\text{IF}[\bm
{r},\bm{r}',t]
    e^{i\left[ S_\text{sys}[\bm{r},t] - S_\text{sys}[\bm{r}',t] \right]} 
    e^{iS_\text{IF}[\bm{r},\bm{r}',t]}
    \right\}
    \label{eq:infinitesimal_rho}
\end{align}
where
\begin{align}
   S_\text{sys}[\bm{r},t]
    &= \int^t_0 dt' 
    \left[ 
    \frac{1}{2} 
    \left[
    1 -3 \Phi(\bm{R})
    \right]
    \mu  \dot{\bm{r}}^2(t') 
    +  
    \left[
    1 + 2\Phi(\bm{R})
    \right]
    \frac{\alpha^2}{4\pi r(t')}  \right],
    \\
     S_\text{IF}[\bm{r},\bm{r}',t]
    &=
    \frac{\lambda^2}{2} \int^t_0 d\tau \int^t_0 d\tau'
    \left[
   r_i (\tau) - r'_i (\tau) 
    \right]
    \mathbb{D}_{ij}(\tau,\tau')
    \left[
    r_j(\tau') + r'_j (\tau')
    \right]
    \nonumber 
    \\
    &\quad + \frac{i\lambda^2}{2} \int^t_0 d\tau \int^t_0 d\tau'\
    \left[ 
    r_i(\tau) - r'_i(\tau) 
    \right]
    \mathbb{N}_{i j}(\tau, \tau')
    \left[ 
   r_j (\tau') - r'_j (\tau') 
    \right],
\end{align}
and $\delta_\epsilon S_\text{IF}$ is defined as
\begin{align}
    \delta_\epsilon S_\text{IF}[r,r',t]
    &\equiv
    i \epsilon \lambda^2 \left[ r_i - r'_j \right]
    \int^t_0 dt' 
    \Bigl(
    \mathbb{N}_{ij}(t,t')
    [r_j(t') - r'_j(t')]
    -
    \frac{i}{2} \mathbb{D}_{ij}(t,t')
    [r_j(t') + r'_j (t')]
    \Bigr).
\end{align}
The first term in Eq.\eqref{eq:infinitesimal_rho} give the unitary evolution term as follows:
\begin{align}
    &
    \int d \bm{r}_{0} \int d \bm{r}_{0}'\rho_\text{s}(\bm{r}_{0},\bm{r}_{0}',0) 
    \int^r_{r_{0}} \mathcal{D}\bm{r} \int^{r'}_{r_{0}'} \mathcal{D}\bm{r}' \
    e^{i\left(S_\text{sys}[\bm{r},t + \epsilon] - S_\text{sys}[\bm{r}',t + \epsilon] \right)} \ e^{iS_\text{IF}[\bm{r},\bm{r}',t]} 
    \notag \\
    &\quad = 
    \int d \bm{r}_{0} \int d \bm{r}_{0}'\rho_\text{s}(\bm{r}_{0},\bm{r}_{0}',0) 
     \notag \\
    & \quad \quad  \times \int^r_{r_{0}} \mathcal{D}\bm{r} \int^{r'}_{r_{0}'} \mathcal{D}\bm{r}'
    \exp
    \left[ 
    i \int^{t + \epsilon}_0 dt' 
    \left\{ 
    \frac{1}{2} \mu_{g} \dot{\bm{r}}^2(t') - V( r(t') ) 
    -
    \frac{1}{2} \mu_{g} \dot{\bm{r}}'^{2}(t') + V( r(t') )  
    \right\} 
    \right] 
    \ e^{iS_\text{IF}[\bm{r},\bm{r}',t]} \notag \\
    &\quad \approx 
    \int 
    \left( 
    \sqrt{\frac{-i \mu_{g}}{2 \pi \epsilon}} 
    \right)^3 
    d\bm{r}_f
    \int 
    \left( 
    \sqrt{\frac{-i \mu_{g}}{2 \pi \epsilon}} 
    \right)^3 
    d\bm{r}'_f
    \exp
    \left[ 
    \frac{i\mu_{g}}{2 \epsilon}(\bm{r} - \bm{r}_f)^2 - i \epsilon V(r_f) 
    - 
    \frac{i\mu_{g}}{2 \epsilon}(\bm{r}' - \bm{r}'_f)^2 + i\epsilon V(r'_f) 
    \right]
     \notag \\
    & \quad \quad \times 
    \ \rho_\text{s}(\bm{r}_f,\bm{r}'_f,t) \notag \\
    &\quad \approx 
   \rho_\text{s}(\bm{r},\bm{r}',t) 
    - 
    i \epsilon 
    \left[ 
    - \frac{1}{2\mu_{g}} \nabla^2_r + V(r) 
    \right] 
   \rho_\text{s}(\bm{r},\bm{r}',t) 
    + 
    i \epsilon 
    \left[ 
    - \frac{1}{2\mu_{g}} \nabla^2_{r'} + V(r') 
    \right] 
   \rho_\text{s}(\bm{r},\bm{r}',t)
    \notag \\
    &\quad = 
   \rho_\text{s}(\bm{r},\bm{r}',t) 
    -
    i \epsilon 
    \langle \bm{r} | 
    \left[ 
  \frac{\hat{\bm{p}}^2}{2\mu_{g}}  + V(\hat{r}), \hat{\rho}_\text{s}(t) \right] 
    |\bm{r}'\rangle ,
    \label{eq:unitary_evolution}
\end{align}
where $\mu_{g}$ and $V(r)$ are
\begin{align}
    \mu_{g} \equiv [1 - 3\Phi(\bm{R})] \mu, 
    \quad
    V(r) \equiv [1 + 2\Phi(\bm{R})] \frac{\alpha^2}{4\pi r}.
    \notag 
\end{align}
We define the following $\bm{X}$ and $\bm{X'}$ as
\begin{align}
    \bm{X}(\bm{r},\bm{r}',t') 
    = \int d r_{0} \int d r_{0}'\rho_\text{s}(\bm{r}_{0},\bm{r}_{0}',0) 
    \int^r_{r_{0}} \mathcal{D}\bm{r} \int^{r'}_{r_{0}'} \mathcal{D}\bm{r}' \
    e^{i\left( S_\text{sys}[r] - S_\text{sys}[r'] + S_\text{IF}[r,r',t'] \right)}
    \ \bm{r}(t'),
    \label{eq:def_X}
    \\
    \bm{X'}(\bm{r},\bm{r}',t') 
    = \int d r_{0} \int d r_{0}'\rho_\text{s}(\bm{r}_{0},\bm{r}_{0}',0) 
    \int^r_{r_{0}} \mathcal{D}\bm{r} \int^{r'}_{r_{0}'} \mathcal{D}\bm{r}' \
    e^{i\left( S_\text{sys}[r] - S_\text{sys}[r'] + S_\text{IF}[r,r',t'] \right)}
    \ \bm{r'}(t'),
    \label{eq:def_X'}
\end{align}
and then the second term may be rewritten as
\begin{align}
    &\int d\bm{r}_{0} \int d\bm{r}_{0}'\,\rho_\text{s}(\bm{r}_{0},\bm{r}_{0}',0) 
    \int_{\bm{r}_0}^{\bm{r}} \mathcal{D}\bm{r} 
     \int_{\bm{r}'_0}^{\bm{r}'} \mathcal{D}\bm{r}'
    \,\,i \delta_\epsilon S_\text{IF}[\bm
{r},\bm{r}',t]
    e^{i\left[ S_\text{sys}[r,t] - S_\text{sys}[r',t] \right]} 
    e^{iS_\text{IF}[r,r',t]}
    \notag
    \\
    &\quad =
    -\epsilon \lambda^2 \left[ r_i - r'_i \right]
    \int^t_0 dt' \
    \Bigl(
    \mathbb{N}_{ij}(t,t')
    [X_j (\bm{r},\bm{r}',t') - X'_j (\bm{r},\bm{r}',t')] 
    -
    \frac{i}{2} \mathbb{D}_{ij}(t,t')
    [X_j (\bm{r},\bm{r}',t') + X'_j (\bm{r},\bm{r},t')]
    \Bigr).
    \label{eq:2nd}
\end{align}
Furthermore, because this term is of second order in $\lambda$, we neglect the factor $e^{iS_\text{IF}[r,r,t]}$, as the influence action becomes relevant only when considering effects beyond $\mathcal{O}(\lambda^2)$.
Therefore, we have the folowing equations
\begin{align}
    X_i (r,r',t') 
    &\approx 
    \int d \bm{r}_{0} \int d \bm{r}_{0}' 
   \rho_\text{s}(\bm{r}_{0},\bm{r}_{0}',0) 
    \int^{\bm{r}}_{\bm{r}_0} \mathcal{D}\bm{r} \int^{\bm{r}'}_{\bm{r}_{0}'} \mathcal{D}\bm{r}' \ e^{i\left(S_\text{sys}[\bm{r},t] - S_\text{sys}[\bm{r}',t]  
    \right)}
    \ r_i (t') \notag \\
    &= 
    \int d\bm{r}_{0} \int d\bm{r}_{0}' 
    \langle \bm{r}| 
    \hat{U}_\text{s}(t,0)
    |\bm{r_{0}}\rangle 
   \rho_\text{s}(\bm{r}_{0},\bm{r}_{0}',0) 
    \langle \bm{r_{0}}'|
    \hat{U}^\dagger_\text{s}(t,0)|\bm{r}'
    \rangle \bm{r}_i (t') 
    \notag \\
    &= 
    \langle \bm{r}| 
    \hat{U}_\text{s}(t,0) \hat{U}^\dagger_\text{s}(t',0) \hat{r}_i \hat{U}_\text{s}(t',0) \hat{\rho}_\text{s}(0) \hat{U}^\dagger_\text{s}(t,0)
    |\bm{r}' \rangle  
    \notag \\
    &= 
    \langle \bm{r}| 
    \hat{U}^\dagger_\text{s}(t' - t,0) \hat{r}_i 
    \hat{U}(t' - t,0) \hat{U}_\text{s}(t,0) \hat{\rho}_\text{s}(0) \hat{U}^\dagger_\text{s}(t,0)
    |\bm{r}' \rangle 
    \notag \\
    &= 
    \langle \bm{r}| 
    \hat{r}_i (t' - t) \rho_\text{s}(t) 
    |\bm{r}'\rangle,
    \label{eq:X^alpha}
\end{align}
and similarly 
\begin{align}
    X'_i (\bm{r},\bm{r}',t)
    &\approx
    \langle \bm{r}| 
    \rho_\text{s}(t) \hat{r}_i (t' - t) 
    |\bm{r}'\rangle.
    \label{eq:X'^alpha}
\end{align}
Using Eqs.\eqref{eq:X^alpha} and \eqref{eq:X'^alpha}, we calculate Eq.\eqref{eq:2nd} as
\begin{align}
   &
    -\epsilon \lambda^2 \left[ r_i - r'_i \right]
    \int^t_0 dt' \
    \Bigl(
    \mathbb{N}_{ij}(t,t')
    [X_j (\bm{r},\bm{r}',t') - X'_j (\bm{r},\bm{r}',t')]
    -
    \frac{i}{2} \mathbb{D}_{ij}(t,t')
    [X_j (\bm{r},\bm{r}',t') + X'_j (\bm{r},\bm{r},t')]
    \Bigr)
    \notag \\
    &\quad =
    -\epsilon \lambda^2 \left[ r_i - r'_i \right]
    \int^t_0 dt' \
    \Bigl(
     \mathbb{N}_{ij}(t,t')
    \langle \bm{r}| 
    \left[ \hat{r}_j (t' - t), \hat{\rho}_\text{s}(t) \right]
    | \bm{r'} \rangle
    -\frac{i}{2} \mathbb{D}_{ij}(t,t')
    \langle \bm{r}| 
    \left\{ \hat{r}_j (t' - t), \hat{\rho}_\text{s}(t) \right\}
    | \bm{r'} \rangle
    \Bigr)
    \notag \\
    &\quad =
    - \epsilon \lambda^2 
    \int^t_0 dt' \
    \Bigl(
    \mathbb{N}_{ij}(t,t')
    \langle \bm{r}| 
    \left[
    \hat{r}_i,
    \left[ \hat{r}_j (t' - t), \hat{\rho}_\text{s}(t) \right]
    \right]
    | \bm{r'} \rangle
    -\frac{i}{2} \mathbb{D}_{ij}(t,t')
    \langle \bm{r}| 
    \left[
    \hat{r}_i,
    \left\{ \hat{r}_j (t' - t), \hat{\rho}_\text{s}(t) \right\}
    \right]
    | \bm{r'} \rangle
    \Bigr).
    \label{eq:N-D_term}
\end{align}
From Eqs.\eqref{eq:unitary_evolution} and \eqref{eq:N-D_term}, 
the infinitesimal time evolution of $\rho_\text{s}(r,r';t)$
can be written as
\begin{align}
    \rho_\text{s}(\bm{r},\bm{r}', t + \epsilon)
    &=\rho_\text{s}(\bm{r},\bm{r}',t) 
    -i \epsilon 
    \langle \bm{r} | 
    \left[ 
  \frac{\hat{\bm{p}}^2}{2\mu_{g}}  + V(\hat{r}), \hat{\rho}_\text{s}(t) \right] 
    |\bm{r}'\rangle
    \notag \\
    &\quad 
    -\epsilon \lambda^2 
    \int^t_0 dt' \
    \Bigl(
    \mathbb{N}_{ij}(t,t')
    \langle \bm{r}| 
    \left[
    \hat{r}_i,
    \left[ \hat{r}_j (t' - t), \hat{\rho}_\text{s}(t) \right]
    \right]
    | \bm{r'} \rangle
    -\frac{i}{2} \mathbb{D}_{ij}(t,t')
    \langle \bm{r}| 
    \left[
    \hat{r}_i,
    \left\{ \hat{r}_j (t' - t), \hat{\rho}_\text{s}(t) \right\}
    \right]
    | \bm{r'} \rangle
    \Bigr).
\end{align}
Up to the first order of the gravitational potential $\Phi$, this is nothing but the QME \eqref{eq:QME_2particle}.

\section{Derivation of Eq.\eqref{eq:QME_Int2}}
\label{app:Derivation_GKSL}
We derive the Markovian QME of the two-level atom, given in \eqref{eq:QME_Int2}. 
In the interaction picture defined by 
$\hat{H}^{(2)}_\text{s}=\Omega_g \hat{\sigma}_z/2$, the QME of the two-level atom, Eq.\eqref{eq:QME_two-level}, is 
\begin{align}
    \frac{\partial}{\partial t} \hat{\rho}^\text{I}_\text{s}(t)
    &=
    \frac{i}{2}  \int^t_0 dt' \ 
    d_i d_j \ \tilde{\mathbb{D}}_{ij}(t,t') 
    \left[ 
    \hat{\sigma}^\text{I}_x (t), 
    \left\{ 
    \hat{\sigma}^\text{I}_x(t'), \hat{\rho}^\text{I}_\text{s}(t) 
    \right\}
    \right]
    - 
    \int^t_0 dt' \ 
    d_i d_j \ \tilde{\mathbb{N}}_{ij}(t,t') 
    \left[ 
    \hat{\sigma}^\text{I}_x (t), 
    \left[
    \hat{\sigma}^\text{I}_x(t'), \hat{\rho}^\text{I}_\text{s}(t) 
    \right]
    \right],
     \label{app:QME}
\end{align}
where 
$\hat{\rho}^\text{I}_\text{s}(t)=e^{i\hat{H}_\text{s} t} \hat{\rho}_\text{s} (t) e^{-i\hat{H}_\text{s} t}$ and 
$\hat{\sigma}^\text{I}_x (t)=e^{i\hat{H}_\text{s} t} \hat{\sigma}_x e^{-i\hat{H}_\text{s} t}$. 
Performing the
variable transformation $t'-t \equiv -\tau$ and changing the integral upper $t$ to $\infty$, we have the Markovian approximated equation, 
\begin{align}
    \frac{\partial}{\partial t} \hat{\rho}^\text{I}_\text{s}(t)
    &=
    \frac{i}{2}  \int^\infty_0 d\tau \ 
    d_i d_j \ \tilde{\mathbb{D}}_{ij}(t,t-\tau) 
    \left[ 
    \hat{\sigma}^\text{I}_x (t), 
    \left\{ 
    \hat{\sigma}^\text{I}_x(t-\tau), \hat{\rho}^\text{I}_\text{s}(t) 
    \right\}
    \right]
    \notag
    \\
    &\quad 
    - 
    \int^\infty_0 d\tau \ 
    d_i d_j \ \tilde{\mathbb{N}}_{ij}(t,t-\tau) 
    \left[ 
    \hat{\sigma}^\text{I}_x (t), 
    \left[
    \hat{\sigma}^\text{I}_x(t-\tau), \hat{\rho}^\text{I}_\text{s}(t) 
    \right]
    \right]. 
     \label{app:QME2}
\end{align}
Substituting 
$\hat{\sigma}^\text{I}_x (t)=e^{i\Omega_g t} \hat{\sigma}_+ + e^{-i\Omega_g t} \hat{\sigma}_{-}$ into the above equation, and ignoring the terms proportional to $e^{\pm i\Omega_g t}$(the rotating-wave approximation), we get  
\begin{align}
    \frac{\partial}{\partial t} \hat{\rho}^\text{I}_\text{s}(t)
    &=
    \frac{i}{2}  \int^\infty_0 d\tau \ 
    d_i d_j \ \tilde{\mathbb{D}}_{ij}(t,t-\tau) 
    \left(e^{i\Omega_g \tau} \left[ 
    \hat{\sigma}_{+}, 
    \left\{ 
    \hat{\sigma}_{-}, \hat{\rho}^\text{I}_\text{s}(t) 
    \right\}
    \right]+e^{-i\Omega_g \tau} \left[ 
    \hat{\sigma}_{-}, 
    \left\{ 
    \hat{\sigma}_{+}, \hat{\rho}^\text{I}_\text{s}(t) 
    \right\}
    \right] \right)
    \notag
    \\
    &\quad 
    - 
    \int^\infty_0 d\tau \ 
    d_i d_j \ \tilde{\mathbb{N}}_{ij}(t,t-\tau) 
   \left(e^{i\Omega_g \tau} \left[ 
    \hat{\sigma}_{+}, 
    \left[ 
    \hat{\sigma}_{-}, \hat{\rho}^\text{I}_\text{s}(t) 
    \right]\right]+e^{-i\Omega_g \tau} \left[ 
    \hat{\sigma}_{-}, 
    \left[
    \hat{\sigma}_{+}, \hat{\rho}^\text{I}_\text{s}(t) 
    \right]
    \right] \right).
     \label{app:QME3}
\end{align}
Here, we define the real quantities $\Gamma_+$, $\Gamma_-$, $S_+$ and $S_-$ as 
\begin{align}
    &\int^{\infty}_0 d\tau d_i d_j \Big( \tilde{\mathbb{N}}_{ij}(t,t-\tau) \pm
    \frac{i}{2} \tilde{\mathbb{D}}_{ij}(t,t-\tau) \Big) e^{i\Omega_g \tau} 
    =\frac{1}{2}\Gamma_{\pm}+iS_{\pm},
     \label{app:GammaS_pm}
\end{align} 
that is, 
\begin{align}
   &\Gamma_{\pm} =2\int^{\infty}_0 d\tau d_i d_j \, \text{Re} \Big[\Big( \tilde{\mathbb{N}}_{ij}(t,t-\tau)\pm
    \frac{i}{2} \tilde{\mathbb{D}}_{ij}(t,t-\tau) \Big) e^{i\Omega_g \tau} \Big] ,
    \label{app:Gamma_pm}
    \\
  &S_{\pm} =\int^{\infty}_0 d\tau d_i d_j \, \text{Im} \Big[\Big( \tilde{\mathbb{N}}_{ij}(t,t-\tau)\pm
    \frac{i}{2} \tilde{\mathbb{D}}_{ij}(t,t-\tau) \Big) e^{i\Omega_g \tau} \Big],
    \label{app:S_pm}
\end{align} 
Using these quantities and noticing that the functions $\tilde{\mathbb{D}}_{ij}(t,t-\tau)$ and $\tilde{\mathbb{N}}_{ij}(t,t-\tau)$ are real, we get the Markovian QME, Eq.\eqref{eq:QME_Int2}, from Eq.\eqref{app:QME} as
\begin{align}
\frac{\partial}{\partial t} \hat{\rho}^\text{I}_\text{s}(t)
&=\Gamma_- \,\hat{\sigma}_{-} \hat{\rho}^\text{I}_\text{s} (t) \hat{\sigma}_{+} +\Gamma_+ \, \hat{\sigma}_{+} \hat{\rho}^\text{I}_\text{s} (t) \hat{\sigma}_{-}
-\Big(\frac{1}{2}\Gamma_- +iS_- \Big)\hat{\sigma}_{+}  \hat{\sigma}_{-} \hat{\rho}^\text{I}_\text{s} (t) -\Big(\frac{1}{2}\Gamma_- -iS_- \Big) \hat{\rho}^\text{I}_\text{s} (t)\hat{\sigma}_{+}  \hat{\sigma}_{-}
\notag
\\
&
\quad -\Big(\frac{1}{2}\Gamma_+ -iS_+ \Big)\hat{\sigma}_{-}  \hat{\sigma}_{+} \hat{\rho}^\text{I}_\text{s} (t) -\Big(\frac{1}{2}\Gamma_+ +iS_+ \Big) \hat{\rho}^\text{I}_\text{s} (t)\hat{\sigma}_{-}  \hat{\sigma}_{+}
\notag
\\
&=-i[\hat{H}_\text{shift}, \hat{\rho}^\text{I}_\text{s}(t)]+\mathcal{D}_-[\hat{\rho}^\text{I}_\text{s}(t)]+\mathcal{D}_+[\hat{\rho}^\text{I}_\text{s}(t)]. 
\label{app:QME4}
\end{align}

\section{Solution of Eq.\eqref{eq:QME_Int2}}
\label{app:solution_QME}
In Sec.~\ref{sec:Case_curve}, we derived the Markovian QME of the two-level atom in the weak gravitational field of interest.
Here, we explicitly get the formal solution of the QME \eqref{eq:QME_Int2}.
Using the eigenstates of the system Hamiltonian $\hat{H}^{(2)}_\text{s}$ as
\begin{align}
    \hat{H}^{(2)}_\text{s} | e \rangle 
    = 
    \frac{\Omega_g}{2} | e \rangle,
    \quad
    \hat{H}^{(2)}_\text{s} | g \rangle
    =
    -\frac{\Omega_g}{2} | g \rangle,
    \label{eq:def_eigenstates_H}
\end{align}
where $|e\rangle$ and $|g\rangle$ denote the excited and ground states, respectively,we can express the ladder operators in this basis as $\hat{\sigma}_+ = | e \rangle \langle g |$ and $\hat{\sigma}_- =  | g \rangle \langle e |$.
Substituting these into Eq.~\eqref{eq:QME_Int2}, the Markovian QME can be expressed in the bra–ket representation as
\begin{align}
   \frac{\partial}{\partial t} \hat{\rho}^\text{I}_\text{s}(t)
   &= 
   \Gamma_- 
   \Big[ | g \rangle \langle e | \hat{\rho}^\text{I}_\text{s}(t) | e \rangle \langle g |
   -
   \frac{1}{2}| e \rangle \langle e | \hat{\rho}^\text{I}_\text{s}(t)
   -
   \frac{1}{2}\hat{\rho}^\text{I}_\text{s}(t) | e \rangle \langle e | \Big]
   \notag
   \\
   &+ 
   \Gamma_+ \Big[ | e \rangle \langle g | \hat{\rho}^\text{I}_\text{s}(t) | g \rangle \langle e |
   -
   \frac{1}{2} | g \rangle \langle g | \hat{\rho}^\text{I}_\text{s}(t)
   -
   \frac{1}{2} \hat{\rho}^\text{I}_\text{s}(t) | g \rangle \langle g | \Big].
   \label{eq:D+_bracket}
\end{align}
From Eq.\eqref{eq:D+_bracket}, the time evolution of the matrix element  $\langle e | \hat{\rho}^\text{I}_\text{s}(t) | e \rangle$ and $\langle e | \hat{\rho}^\text{I}_\text{s}(t) | g \rangle$ is obtained as
\begin{align}
    \frac{\partial}{\partial t} 
    \langle e | \hat{\rho}^\text{I}_\text{s}(t) | e \rangle
    &=
    - \Gamma_- \langle e | \hat{\rho}^\text{I}_\text{s}(t) | e \rangle 
    + \Gamma_+  \langle g | \hat{\rho}^\text{I}_\text{s}(t) | g \rangle,
    \label{eq:QME_element1}
    \\
    \frac{\partial}{\partial t} 
    \langle e | \hat{\rho}^\text{I}_\text{s}(t) | g \rangle
    &= -\frac{1}{2} (\Gamma_+ +\Gamma_-) \langle e | \hat{\rho}^\text{I}_\text{s}(t) | g \rangle.
    \label{eq:QME_element2}
\end{align}
It is easy to solve Eq.\eqref{eq:QME_element2}, and the solution is 
\begin{align} 
    \langle e | \hat{\rho}^\text{I}_\text{s}(t) | g \rangle
    &= e^{-\frac{1}{2} (\Gamma_+ +\Gamma_-)t
}\langle e | \hat{\rho}^\text{I}_\text{s}(0) | g \rangle, 
    \label{eq:Sol_eg}
\end{align}
and, because 
$\langle g|\hat{\rho}^\text{I}_\text{s}(t)|e\rangle$ is given by the complex conjugate of $\langle e|\hat{\rho}^\text{I}_\text{s}(t)|g\rangle$, we also have
\begin{align} 
    \langle g | \hat{\rho}^\text{I}_\text{s}(t) | e \rangle
    &= \langle e | \hat{\rho}^\text{I}_\text{s}(t) | g\rangle^* 
\notag
\\
    &= e^{-\frac{1}{2} (\Gamma_+ +\Gamma_-)t
}\langle g | \hat{\rho}^\text{I}_\text{s}(0) | e \rangle. 
    \label{eq:Sol_ge}
\end{align}
Because the trace of the density matrix, $\hat{\rho}^\text{I}_\text{s}(t)$, is unity, we have  $\langle e | \hat{\rho}^\text{I}_\text{s}(t) | e \rangle+\langle g | \hat{\rho}^\text{I}_\text{s}(t) | g \rangle=1$, and then Eq.\eqref{eq:QME_element1} is written as 
\begin{align}
    \frac{\partial}{\partial t} 
    \langle e | \hat{\rho}^\text{I}_\text{s}(t) | e \rangle
    &=
    - \Gamma_- \langle e | \hat{\rho}^\text{I}_\text{s}(t) | e \rangle 
    + \Gamma_+  (1-\langle e | \hat{\rho}^\text{I}_\text{s}(t) | e \rangle) 
\end{align}
The solution of this equation is given as 
\begin{align}
    \langle e | \hat{\rho}^\text{I}_\text{s}(t) | e \rangle
    &=e^{-(\Gamma_+ + \Gamma_-)t} \langle e | \hat{\rho}^\text{I}_\text{s}(0) | e \rangle +\frac{\Gamma_+}{\Gamma_+ +\Gamma_-}(1-e^{-(\Gamma_+ + \Gamma_-)t}), 
\end{align}
and hence
\begin{align}
    \langle g | \hat{\rho}^\text{I}_\text{s}(t) | g \rangle
    &=1-\langle e | \hat{\rho}^\text{I}_\text{s}(t) | e \rangle
    \notag
    \\
    &=1-e^{-(\Gamma_+ + \Gamma_-)t} \langle e | \hat{\rho}^\text{I}_\text{s}(0) | e \rangle -\frac{\Gamma_+}{\Gamma_+ +\Gamma_-}(1-e^{-(\Gamma_+ + \Gamma_-)t})
    \notag
    \\
    &=e^{-(\Gamma_+ + \Gamma_-)t} \langle g | \hat{\rho}^\text{I}_\text{s}(0) | g \rangle +\frac{\Gamma_-}{\Gamma_+ +\Gamma_-}(1-e^{-(\Gamma_+ + \Gamma_-)t}).
\end{align}
Therefore the formal solution of the Markovian QME is 
\begin{align}
    \hat{\rho}^\text{I}_\text{s}(t) &=
        \begin{pmatrix}
        [\langle e | \hat{\rho}^\text{I}_\text{s}(0) | e \rangle - a_s] e^{- \Gamma t} + a_s & \langle e | \hat{\rho}^\text{I}_\text{s}(0) | g \rangle \ e^{- \frac{1}{2} \Gamma  t} \\
        \langle g | \hat{\rho}^\text{I}_\text{s}(0) | e \rangle \ e^{- \frac{1}{2} \Gamma t} & e^{-\Gamma t} \langle g | \hat{\rho}^\text{I}_\text{s}(0) | g \rangle +(1-a_s)(1-e^{-\Gamma t})
    \end{pmatrix},
    \label{eq:formal_sol}
\end{align}
where $\Gamma=\Gamma_+ +\Gamma_-$ and $a_s =\Gamma_+/(\Gamma_+ +\Gamma_-)$. 
When the initial condition of $\hat{\rho}^\text{I}_\text{s}$ is yielded by $\langle e|\hat{\rho}^\text{I}_\text{s}(0)|e\rangle=1$, $\langle e|\hat{\rho}^\text{I}_\text{s}(0)|g\rangle=0= \langle g|\hat{\rho}^\text{I}_\text{s}(0)|e\rangle=\langle g|\hat{\rho}^\text{I}_\text{s}(0)|g\rangle$, we reproduce the solution given in Eq.\eqref{eq:solve_QME}.

\section{Analysis of $\mathbb{D}$ and $\mathbb{N}$}
\label{app:Analysys_two-point}
In this section, we drive the dissipation and noise kernels $\mathbb{D}(x,x')$ and $\mathbb{N}(x,x')$ defined by Eq.\eqref{eq:def_DN}.
These functions can be obtained through the Feynmann propagator $G_\text{F}(x,x')$, so we derive it in a weak gravitational field.
To this end, we start from computing the propagator $G^0_\text{F}(x,x')$ in flat spacetime.
The propagator $G_\text{F}(x,x')$ satisfies the following differential equation:
\begin{align}
    \eta^{\mu\nu} \partial^x_\mu \partial^x_\nu \, G^0_\text{F}(x,x') 
    = - \delta^4(x-x'),
    \quad
    G^0_\text{F}(x,x')
    \equiv
    i \langle T[\phi(x)\phi(x')]\rangle,
    \label{eq:def_Fey_pro}
\end{align}
where $T$ denotes the time-ordered product.
We are interested in the propagator of the thermal scalar field, so we solve Eq.\eqref{eq:def_Fey_pro} by imaginary time method.
When we introduce a new parameter $\tau \equiv -i t$, the above differential equation can be rewritten as
\begin{align}
    ( \partial^2_\tau + \nabla^2_x ) \, G^0_\text{E}(x,x') 
    = -i \delta(\tau - \tau') \delta^3(\bm{x} - \bm{x}'),
    \label{eq:Fey_pro_Im}
\end{align}
where the label E means the function is described by a imaginary time.
The solution of Eq.\eqref{eq:Fey_pro_Im} can be given as
\begin{align}
    G^0_\text{E}(x,x') 
    &=
    \frac{i}{\beta} \sum_{n = -\infty}^\infty \int \frac{d^3p}{(2\pi)^3} 
    \frac{ e^{-i \omega_n (\tau - \tau')} }{\omega^2_n + |\bm{p}|^2} \,
    e^{i \bm{p} \cdot (\bm{x} - \bm{x}')}
    \notag 
    \\
    &=
    - \frac{i}{\beta} \sum_{n = -\infty}^\infty \int \frac{d^4p}{(2\pi)^4}
    \, \frac{\rho(p^0,\bm{p}) }{i \omega_n - p^0} e^{-i \omega_n (\tau - \tau')}
    e^{i \bm{p} \cdot (\bm{x} - \bm{x}')},
    \label{eq:sol_Fey}
\end{align}
where $\omega_n = 2 \pi n/\beta$,  the parameter $\beta$ is an inverse temperature, and the spectral density
\begin{align}
    \rho(p^0,\bm{p}) 
    &\equiv
    2 \pi \left[ \theta(p^0) - \theta( - p^0) \right] 
    \delta\left[ -(p^0)^2 + |\bm{p}|^2 \right].
    \label{eq:def_spectle}
\end{align}
Furthermore, because the following relation
\begin{align}
    \frac{1}{\beta} \sum^\infty_{n = - \infty} \frac{ e^{-i \omega_n (\tau - \tau')} }{ i \omega_n - p^0}
    =
    -
    \left[
    \theta(\tau - \tau') + n_B(p^0)
    \right]
    e^{- p^0 (\tau - \tau')}
    \notag
\end{align}
holds \cite{Tremblay_2008,Nieto_1995}, the solution \eqref{eq:sol_Fey} may be written as
\begin{align}
    G^0_\text{E}(x,x') 
    &=
    - \frac{i}{\beta} \sum_{n = -\infty}^\infty \int \frac{d^4p}{(2\pi)^4}
    \, \frac{\rho(p^0,\bm{p}) }{i \omega_n - p^0} e^{-i \omega_n (\tau - \tau')}
    e^{i \bm{p} \cdot (\bm{x} - \bm{x}')}
    \notag \\
    &=
    i \int \frac{d^4p}{(2\pi)^4} \, \rho(p^0,\bm{p}) \, 
    e^{ - p^0 (\tau - \tau') + i \bm{p} \cdot ( \bm{x} - \bm{x'}) }
    \left[ \theta( \tau - \tau' ) + n_B(p^0) \right].
    \label{eq:sol_Fey2}
\end{align}
Because of \eqref{eq:sol_Fey2}, the Feynmann propagator $G_\text{F}(x,x')$ can be obtained by replacing $\tau - \tau'$ with $i(t - t')$ in Eq.\eqref{eq:sol_Fey2}\cite{Bellac_1996,Lundberg_2021,Dolan-Jackiw_1974}.
\begin{align}
    G^0_\text{F}(x,x')
    &=
    i \int \frac{d^4p}{(2\pi)^4} \, \rho(p^0,\bm{p}) \, 
    e^{ i p^\mu (x - x')_\mu }
    \left[ \theta( t - t' ) + n_B(p^0) \right]
    \notag 
    \\
    &=
    i \int \frac{d^3p}{(2\pi)^3} \frac{1}{ 2|\bm{p|} } 
    \left[
    \theta(t - t') e^{ -i |\bm{p}|(t - t') + i \bm{p} \cdot ( \bm{x} - \bm{x'} )}
    +
    \theta(t' - t) e^{ i |\bm{p}|(t - t') - i \bm{p} \cdot ( \bm{x} - \bm{x'} )}
    \right.
    \notag \\
    &\hspace{5cm} \left.
    + 
    n_B(|\bm{p}|) 
    \left\{
    e^{ -i |\bm{p}|(t - t') + i \bm{p} \cdot ( \bm{x} - \bm{x'} )}
    +
    e^{ i |\bm{p}|(t - t') - i \bm{p} \cdot ( \bm{x} - \bm{x'} )}
    \right\}
    \right]
    \notag 
    \\
    &=
    \int \frac{d^4 p}{(2\pi)^4} \, e^{ i p^\mu (x - x')_\mu } 
    \left[
    \frac{1}{ p^2 - i \epsilon } + 2 \pi i \, n_B(p^0) \, \delta(p^2)
    \right]
    \label{eq:sol_Fey3}
\end{align}

Next, we derive the Feynmann propagator $G_\text{F}(x,x')$ in a weak gravitational field.
When we assume that the solution $G_\text{F}(x,x')$ is perturbatively given by $G_\text{F}(x,x') \approx G^0_\text{F}(x,x') + \delta G_\text{F}(x,x')$ with respect to the gravitational potential $\Phi$, the perturbative solution $\delta G_\text{F}(x,x')$ satisfies the following differential equation:
\begin{align}
    \eta^{\mu\nu}\partial^x_\mu \partial^x_\nu \, \delta G_\text{F}(x,x') 
    \approx -4 \Phi(\bm{x}) \partial^2_t G^0_\text{F}(x,x')
    \notag
\end{align}
Note that the derivation from now on is valid for a wavelength scale of scalar field larger than the gravitational radius of the source.
We solve this equation by the imaginary time method and obtain the solution in real time by analytic continuation.
In an imaginary time, the differential equation is written as
\begin{align}
    ( \partial^2_\tau + \nabla^2_x ) \, \delta G_\text{E}(x,x')
    =
    4\Phi(\bm{x}) \partial^2_\tau G^0_\text{E}(x,x'),
    \label{eq:def_delta_Fey}
\end{align}
so this solution is formally represented as
\begin{align}
    \delta G_\text{E}(x,x')
    = i \int^\beta_0 d\tau_y \int d^3y \, 4\Phi(\bm{y}) G^0_\text{E}(x,y) 
    \partial^2_{\tau_y} G^0_\text{E}(y,x')
    \label{eq:sol_delta_Fey}
\end{align}
By using the expression of $G^0_\text{E}(x,x')$ such as Eqs.\eqref{eq:sol_Fey} and \eqref{eq:sol_Fey2}, Eq.\eqref{eq:sol_delta_Fey} can be rewritten as
\begin{align}
    &\delta G_\text{E}(x,x')
    \notag 
    \\
    &=
    \frac{i}{\beta^2} \int \frac{d^3p}{ (2\pi)^3 } \int \frac{d^3k}{ (2\pi)^3 }
    \, e^{ i \bm{p} \cdot \bm{x} - i \bm{k} \cdot \bm{x}' }
    \sum^\infty_{n = - \infty} \frac{ e^{ -i \omega_n \tau } }{ \omega^2_n + |\bm{p}|^2 }
    \sum^\infty_{m = - \infty} \frac{ \omega^2_m e^{ i \omega_m \tau' } }{ \omega^2_m + |\bm{p}|^2 }
    \notag 
    \\
    &\quad \times \int^\beta_0 d\tau_y \, e^{ i ( \omega_n - \omega_m) \tau_y }
    \int d^3y \, 4\Phi(\bm{y}) \, e^{-i ( \bm{p} - \bm{k} ) \cdot \bm{y} }
    \notag 
    \\
    &=
    \frac{i}{\beta} \int \frac{d^3p}{ (2\pi)^3 } \int \frac{d^3k}{ (2\pi)^3 }
    \, e^{ i \bm{p} \cdot \bm{x} - i \bm{k} \cdot \bm{x}' }
    \sum^\infty_{n = - \infty} 
    \frac{ \omega^2_n e^{ -i \omega_n (\tau-\tau') } }{ ( \omega^2_n + |\bm{p}|^2 )( \omega^2_n + |\bm{k}|^2 ) }
    \left[
    - \frac{16 \pi GM}{|\bm{p} - \bm{k}|^2}
    \right]
    \notag 
    \\
    &=
    - \frac{ 16 \pi i GM }{ \beta }
    \int \frac{d^3p}{ (2\pi)^3 } \int \frac{d^3k}{ (2\pi)^3 }
    \frac{ e^{ i \bm{p} \cdot \bm{x} - i \bm{k} \cdot \bm{x}' } }
    { |\bm{p} - \bm{k}|^2 ( |\bm{p}|^2 - |\bm{k}|^2 ) }
    \sum^\infty_{n = - \infty} 
    \left[
    |\bm{p}|^2 
    \frac{ e^{ -i \omega_n (\tau-\tau') } }{  \omega^2_n + |\bm{p}|^2  }
    -
    |\bm{k}|^2
    \frac{ e^{ -i \omega_n (\tau-\tau') } }{  \omega^2_n + |\bm{k}|^2  }
    \right]
    \notag 
    \\
    &=
    - \frac{ 16 \pi i GM }{ \beta }
    \int \frac{d^3p}{ (2\pi)^3 } \int \frac{d^3k}{ (2\pi)^3 }
    \frac{ e^{ i \bm{p} \cdot \bm{x} - i \bm{k} \cdot \bm{x}' } }
    { |\bm{p} - \bm{k}|^2 ( |\bm{p}|^2 - |\bm{k}|^2 ) }
    \notag 
    \\
    &\quad 
    \times 
    \left[
    -
    |\bm{p}|^2 
    \sum^\infty_{n = - \infty} 
    \int \frac{dp^0}{2\pi} 
    \frac{ \rho( p^0, \bm{p} ) }{ i \omega_n - p^0 } \, e^{ -i \omega_n (\tau-\tau') }
    +
    |\bm{k}|^2
    \sum^\infty_{n = - \infty}  
    \int \frac{dk^0}{2\pi} 
    \frac{ \rho( k^0, \bm{k} ) }{ i \omega_n - k^0 } \, e^{ -i \omega_n (\tau-\tau')}
    \right]
    \notag 
    \\
    &=
    - 16 \pi i GM 
    \int \frac{d^3p}{ (2\pi)^3 } \int \frac{d^3k}{ (2\pi)^3 }
    \frac{ e^{ i \bm{p} \cdot \bm{x} - i \bm{k} \cdot \bm{x}' } }
    { |\bm{p} - \bm{k}|^2 ( |\bm{p}|^2 - |\bm{k}|^2 ) }
    \left[
    |\bm{p}|^2 \int \frac{ dp^0 }{ 2\pi } \rho(p^0,\bm{p}) 
    [ \theta(\tau - \tau') + n_B(p^0) ]
    e^{- p^0 (\tau - \tau') }
    \right.
    \notag \\
    & \quad 
    \left.
    -
    |\bm{k}|^2 \int \frac{ dk^0 }{ 2\pi } \rho(k^0,\bm{k}) 
    [ \theta(\tau - \tau') + n_B(k^0) ]
    e^{- k^0 (\tau - \tau') }
    \right]
    \notag 
    \\
    &=
    - 16 \pi i GM
    \int \frac{d^4p}{ (2\pi)^4 } \int \frac{d^3k}{ (2\pi)^3 }
    \frac{ |\bm{p}|^2 \rho(p^0, \bm{p})e^{ - p^0 (\tau - \tau')} }{ |\bm{p} - \bm{k}|^2 ( |\bm{p}|^2 - |\bm{k}|^2 ) }
    \, [ \theta(\tau - \tau') + n_B(p^0) ]
    \, 
    \left[ 
    e^{ i \bm{p} \cdot \bm{x} - i \bm{k} \cdot \bm{x}' }
    +
    e^{ i \bm{k} \cdot \bm{x} - i \bm{p} \cdot \bm{x}' }
    \right],
    \notag
\end{align}
so the solution $\delta G_\text{F}(x,x')$ of Eq.\eqref{eq:def_delta_Fey} can be given as
\begin{align}
    \delta G_\text{F}(x,x')
    &=
    - 16 \pi i GM
    \int \frac{d^4p}{ (2\pi)^4 } \int \frac{d^3k}{ (2\pi)^3 }
    \frac{ |\bm{p}|^2 \rho(p^0, \bm{p}) e^{ - i p^0 (t - t')} }
    { |\bm{p} - \bm{k}|^2 ( |\bm{p}|^2 - |\bm{k}|^2 ) }
    \notag 
    \\
    & \quad
    \times
    \, 
    [ \theta(t - t') + n_B(p^0) ]
    \left[ 
    e^{ i \bm{p} \cdot \bm{x} - i \bm{k} \cdot \bm{x}' }
    +
    e^{ i \bm{k} \cdot \bm{x} - i \bm{p} \cdot \bm{x}' }
    \right].
    \label{eq:sol_delta_Fey2}
\end{align}

Because of Eqs.\eqref{eq:sol_Fey3} and \eqref{eq:sol_delta_Fey2}, the dissipation kernel $\mathbb{D}(x,x')$ and the noise kernel $\mathbb{N}(x,x')$ in a weak gravitational field can be obtained.
Based on Ref.\cite{Christensen_1976,Hsiang_2024}, there is a following relation between Feynmann propagator $G_\text{F}(x,x')$ and these functions:
\begin{align}
    G_\text{F}(x,x') 
    =
    \frac{1}{2} 
    \left(
    \mathbb{D}(x,x') + G_\text{adv}(x,x')
    \right)
    +
    i \, \mathbb{N}(x,x'),
    \label{eq:Relation_Fey-D&N}
\end{align}
where $G_\text{adv}(x,x')$ is the advanced Green function. 
Therefore, Eq.\eqref{eq:Relation_Fey-D&N} means that the kernels $\mathbb{D}$ and $\mathbb{N}$ are obtained as $\mathbb{D}(x,x')=2\text{Re}G_\text{F}(x,x') \theta(t-t')$ and $\mathbb{N}(x,x')=\text{Im}G_\text{F}(x,x')$, respectively.
Explicitly, the kernels are 
\begin{align}
    \mathbb{D}(x,x')
    &=2\text{Re}G_\text{F}(x,x') \theta(t-t')
    \notag \\
    &=
    \left[ G_\text{F}(x,x') + G^\ast_\text{F}(x,x') \right]\theta(t-t')
    \notag 
    \\
    &=
    \left[ G^0_\text{F}(x,x') + G^{0\ast}_\text{F}(x,x') \right]\theta(t-t')
    +
    \left[ \delta G_\text{F}(x,x') + \delta G^\ast_\text{F}(x,x') \right]\theta(t-t')
    \notag \\
    &=
    i \, \theta(t - t') \int \frac{d^3p}{(2\pi)^3} \, \frac{1}{2|\bm{p}|}
    I^{-}_p(t,t')e^{i \bm{p} \cdot (\bm{x} - \bm{x'})}
    \notag \\
    & \quad 
    - 8 \pi i GM \, \theta(t - t')
    \int \frac{d^3p}{(2\pi)^3} \int \frac{d^3k}{(2\pi)^3}
    \frac{
    e^{i \bm{p} \cdot \bm{x} }
    e^{- i \bm{k} \cdot \bm{x}'}
    }
    {
    |\bm{p} - \bm{k}|^2 (|\bm{p}|^2 - |\bm{k}|^2)
    }
    \left[
    |\bm{p}| I^{-}_p(t , t') - |\bm{k}| I^{-}_k(t , t')
    \right],
\end{align}
and 
\begin{align}
    \mathbb{N}(x,x')&=
    \text{Im} G_\text{F}(x,x')
    \notag 
    \\
    &=
    \frac{1}{2i} \left[ G_\text{F}(x,x') - G^\ast_\text{F}(x,x') \right]
    \notag 
    \\
    &=
    \frac{1}{2i} \left[ G^0_\text{F}(x,x') - G^{0\ast}_\text{F}(x,x') \right]
    +
    \frac{1}{2i} \left[ \delta G_\text{F}(x,x') - \delta G^\ast_\text{F}(x,x') \right]
    \notag 
    \\
    &=
    \frac{1}{2} \int \frac{d^3p}{(2\pi)^3} \, \frac{1}{2|\bm{p}|}
    I^{+}_p(t,t') \,
    e^{i \bm{p} \cdot (\bm{x} - \bm{x'})} 
    \notag 
    \\
    & \quad 
    -
    4 \pi GM
    \int \frac{d^3p}{(2\pi)^3} \int \frac{d^3k}{(2\pi)^3}
    \frac{
    e^{i \bm{p} \cdot \bm{x} }
    e^{- i \bm{k} \cdot \bm{x}'}
    }
    {
    |\bm{p} - \bm{k}|^2 (|\bm{p}|^2 - |\bm{k}|^2)
    }
    \left[
    |\bm{p}| I^{+}_p(t , t') - |\bm{k}| I^{+}_k(t , t')
    \right],
\end{align}
so the dissipation and noise kernels $\mathbb{D}(x,x')=
    \mathbb{D}_0(x,x') + \delta \mathbb{D}(x,x')$ and $\mathbb{N}(x,x')=
    \mathbb{N}_0 (x,x') + \delta \mathbb{N} (t,\bm{x},x')$ are given by
\begin{align}
    \mathbb{D}_0(x,x')
    &=
    i \, \theta(t - t') \int \frac{d^3p}{(2\pi)^3} \, \frac{1}{2|\bm{p}|}
    I^{-}_p(t,t') \, e^{i \bm{p} \cdot (\bm{x} - \bm{x'})},
    \notag 
    \\
    \mathbb{N}_0 (x,x')
    &=
    \frac{1}{2} \int \frac{d^3p}{(2\pi)^3} \, \frac{1}{2|\bm{p}|}
    I^{+}_p(t,t') \,
    e^{i \bm{p} \cdot (\bm{x} - \bm{x'})}, 
    \notag \\
    \delta \mathbb{D}(x,x')
    &=
    - 8 \pi i GM \, \theta(t - t')
    \int \frac{d^3p}{(2\pi)^3} \int \frac{d^3k}{(2\pi)^3}
    \frac{
    e^{i \bm{p} \cdot \bm{x} }
    e^{- i \bm{k} \cdot \bm{x}'}
    }
    {
    |\bm{p} - \bm{k}|^2 (|\bm{p}|^2 - |\bm{k}|^2)
    }
    \left[
    |\bm{p}| I^{-}_p(t , t') - |\bm{k}| I^{-}_k(t , t')
    \right]
    \notag 
    \\
    \delta \mathbb{N} (x,x')
    &=
    -
    4 \pi GM
    \int \frac{d^3p}{(2\pi)^3} \int \frac{d^3k}{(2\pi)^3}
    \frac{
    e^{i \bm{p} \cdot \bm{x} }
    e^{- i \bm{k} \cdot \bm{x}'}
    }
    {
    |\bm{p} - \bm{k}|^2 (|\bm{p}|^2 - |\bm{k}|^2)
    }
    \left[
    |\bm{p}| I^{+}_p(t , t') - |\bm{k}| I^{+}_k(t , t')
    \right].
    \notag
\end{align}
These functions are nothing but ones defined by Eqs.\eqref{eq:def_D0}, \eqref{eq:def_deltaD}, \eqref{eq:def_N0}, and \eqref{eq:def_deltaN}.

\section{Calculation of $\Gamma_{\pm}$ and $\gamma_g$}
\label{app:detail_Ag-K}
In this section, we compute $\Gamma_{\pm}$ and $\gamma_g$ given in Eqs.\eqref{eq:Gamma_pm} and \eqref{eq:gamma_g}.  
To do this, we need the explicit forms of $\tilde{D}_{ij}$ and $\tilde{N}_{ij}$. 
Substituting Eqs.\eqref{eq:def_D0}, \eqref{eq:def_deltaD}, \eqref{eq:def_N0}, and \eqref{eq:def_deltaN} into \eqref{eq:tilD} and \eqref{eq:tilN}, we get 
\begin{align}
    \tilde{\mathbb{D}}_{ij}(t,t')
    &=
    [ 1 + 4\Phi(\bm{R}) ]
    i\theta(t - t')
    \int \frac{d^3p}{(2\pi)^3} \frac{p_i p_j}{2|\bm{p}|}
     I^{-}_p(t,t')
    \nonumber 
    \\
    & \quad  +8 \pi i R \Phi (\bm{R}) \theta(t - t')
    \int \frac{d^3p}{(2\pi)^3} \int \frac{d^3k}{(2\pi)^3}
    \frac{p_i k_j e^{ i (\bm{p}-\bm{k}) \cdot \bm{R} }}
    { |\bm{p} - \bm{k}|^2 ( |\bm{p}|^2 - |\bm{k}|^2) }
    \left[
    |\bm{p}| I^{-}_p(t,t') - |\bm{k}| I^{-}_k(t,t')
    \right],
    \label{app:tilD}
    \\
    \tilde{\mathbb{N}}_{ij}(t,t')
    & =[ 1 + 4\Phi(\bm{R}) ]
    \frac{1}{2}
    \int \frac{d^3p}{(2\pi)^3} \frac{p_i p_j}{2|\bm{p}|}
     I^{+}_p(t,t')
    \nonumber 
    \\
    & \quad  +4 \pi R \Phi (\bm{R}) 
    \int \frac{d^3p}{(2\pi)^3} \int \frac{d^3k}{(2\pi)^3}
    \frac{p_i k_j e^{ i (\bm{p}-\bm{k}) \cdot \bm{R} }}
    { |\bm{p} - \bm{k}|^2 ( |\bm{p}|^2 - |\bm{k}|^2) }
    \left[
    |\bm{p}| I^{+}_p(t,t') - |\bm{k}| I^{+}_k(t,t')
    \right]. 
    \label{app:tilN}
\end{align}
Since the relations, 
\begin{align}
    &\int^\infty_0 d\tau \, \text{Re}[\theta(\tau)
     I^{-}_p(t,t-\tau)e^{i\Omega_g \tau}]
     =\pi\delta(|\bm{p}|-\Omega_g),
    \label{app:rel1}
    \\
     &\int^\infty_0 d\tau \, \text{Re}[
     I^{+}_p(t,t-\tau)e^{i\Omega_g \tau}]
     =\pi(2n_B (\Omega_g)+1)\delta(|\bm{p}|-\Omega_g), 
\end{align}
hold, we can perform the $\tau$ integral of $\Gamma_{\pm}$. 
Explicitly, the quantities $\Gamma_\pm$ are 
\begin{align}
    \Gamma_\pm &=2\int^{\infty}_0 d\tau d_i d_j \, \text{Re} \Big[\Big( \tilde{\mathbb{N}}_{ij}(t,t-\tau)\pm
    \frac{i}{2} \tilde{\mathbb{D}}_{ij}(t,t-\tau) \Big) e^{i\Omega_g \tau} \Big]
    \nonumber 
    \\
    &=2[2n_B (\Omega_g)+1]\Big\{[1+4\Phi(\bm{R})] \frac{\pi}{2}
    \int \frac{d^3p}{(2\pi)^3} \frac{(\bm{p} \cdot \bm{d})^2}{2|\bm{p}|} \delta(|\bm{p}|-\Omega_g)
    \nonumber 
    \\
    &\quad
     +4 \pi^2 R \Phi (\bm{R}) 
    \int \frac{d^3p}{(2\pi)^3} \int \frac{d^3k}{(2\pi)^3}
    \frac{(\bm{p}\cdot \bm{d}) (\bm{k}\cdot \bm{d}) e^{ i (\bm{p}-\bm{k}) \cdot \bm{R} }}
    { |\bm{p} - \bm{k}|^2 ( |\bm{p}|^2 - |\bm{k}|^2) }
    \left[
    |\bm{p}| \delta(|\bm{p}|-\Omega_g) - |\bm{k}| \delta(|\bm{k}|-\Omega_g)
    \right] \Big \}
    \nonumber 
    \\
     &\quad \pm 2\Big\{[1+4\Phi(\bm{R})] \frac{\pi}{2}
    \int \frac{d^3p}{(2\pi)^3} \frac{(\bm{p} \cdot \bm{d})^2}{2|\bm{p}|} \delta(|\bm{p}|-\Omega_g)
    \nonumber 
    \\
    &\quad
     +4 \pi^2 R \Phi (\bm{R}) 
    \int \frac{d^3p}{(2\pi)^3} \int \frac{d^3k}{(2\pi)^3}
    \frac{(\bm{p}\cdot \bm{d}) (\bm{k}\cdot \bm{d}) e^{ i (\bm{p}-\bm{k}) \cdot \bm{R} }}
    { |\bm{p} - \bm{k}|^2 ( |\bm{p}|^2 - |\bm{k}|^2) }
    \left[
    |\bm{p}| \delta(|\bm{p}|-\Omega_g) - |\bm{k}| \delta(|\bm{k}|-\Omega_g)
    \right]\Big \}
    \nonumber 
    \\
    &
    =
     \begin{cases}
        n_B (\Omega_g) \gamma_g
        &  (+) \\[2pt]
        [n_B (\Omega_g)+1] \gamma_g
        & (-)
    \end{cases}
    ,
    \label{app:Gammas}
\end{align}
where $\gamma_g$ is given as 
\begin{align}
    \gamma_g 
    &=\pi [1+4\Phi(\bm{R})]
    \int \frac{d^3p}{(2\pi)^3} \frac{(\bm{p} \cdot \bm{d})^2}{|\bm{p}|}   \delta(|\bm{p}|-\Omega_g)
    \nonumber 
    \\
    &\quad
     +16 \pi^2 R \Phi (\bm{R}) 
    \int \frac{d^3p}{(2\pi)^3} \int \frac{d^3k}{(2\pi)^3}
    \frac{(\bm{p}\cdot \bm{d}) (\bm{k}\cdot \bm{d}) e^{ i (\bm{p}-\bm{k}) \cdot \bm{R} }}
    { |\bm{p} - \bm{k}|^2 ( |\bm{p}|^2 - |\bm{k}|^2) }
    \left[
    |\bm{p}| \delta(|\bm{p}|-\Omega_g) - |\bm{k}| \delta(|\bm{k}|-\Omega_g)
    \right] \Big \}. 
    \label{app:gamma_g}
\end{align}
All we have to do is to compute this $\gamma_g$. It is easy to compute the first term of $\gamma_g$: 
\begin{align}
  &\pi [1+4\Phi(\bm{R})]
    \int \frac{d^3p}{(2\pi)^3} \frac{(\bm{p} \cdot \bm{d})^2}{|\bm{p}|}   \delta(|\bm{p}|-\Omega_g)
    \nonumber 
    \\
    &=\pi [1+4\Phi(\bm{R})]\frac{1}{(2\pi)^3}
    \int^\infty_0 dp \,p^2 \int^\pi_0 d\theta \sin \theta \int^{2\pi}_0 d\varphi \frac{p^2 d^2 \cos^2\theta}{p}   \delta(p-\Omega_g)
    \nonumber 
    \\ 
    &=[1+4\Phi(\bm{R})]\frac{\Omega^3_g d^2}{6\pi}. 
    \label{app:1st_g}
\end{align}
The second term of $\gamma_g$ is written as 
\begin{align}
   &16 \pi^2 R \Phi (\bm{R}) 
    \int \frac{d^3p}{(2\pi)^3} \int \frac{d^3k}{(2\pi)^3}
    \frac{(\bm{p}\cdot \bm{d}) (\bm{k}\cdot \bm{d}) e^{ i (\bm{p}-\bm{k}) \cdot \bm{R} }}
    { |\bm{p} - \bm{k}|^2 ( |\bm{p}|^2 - |\bm{k}|^2) }
    \left[
    |\bm{p}| \delta(|\bm{p}|-\Omega_g) - |\bm{k}| \delta(|\bm{k}|-\Omega_g)
    \right] 
    \nonumber
    \\
     &\quad =32 \pi^2 R \Phi (\bm{R})  
    \int \frac{d^3p}{(2\pi)^3} \int \frac{d^3k}{(2\pi)^3}
    \frac{(\bm{p}\cdot \bm{d}) (\bm{k}\cdot \bm{d}) e^{ i (\bm{p}-\bm{k}) \cdot \bm{R} }}
    { |\bm{p} - \bm{k}|^2 ( |\bm{p}|^2 - |\bm{k}|^2) }
    |\bm{p}| \delta(|\bm{p}|-\Omega_g)
    \nonumber
    \\
     &\quad =32 \pi^2 R \Phi (\bm{R}) \Omega_g
    \int \frac{d^3p}{(2\pi)^3} \int \frac{d^3k}{(2\pi)^3}
    \frac{(\bm{p}\cdot \bm{d}) (\bm{k}\cdot \bm{d}) e^{ i (\bm{p}-\bm{k}) \cdot \bm{R} }}
    { |\bm{p} - \bm{k}|^2 ( \Omega_g^2 - |\bm{k}|^2) }
    \delta(|\bm{p}|-\Omega_g)
     \nonumber
    \\
     &\quad =8 \pi R \Phi (\bm{R}) \Omega_g \int \frac{d^3y}{|\bm{y}|} \Big(\int \frac{d^3p}{(2\pi)^3} \bm{p}\cdot \bm{d} \, e^{-i\bm{p} \cdot (\bm{y}-\bm{R})}  \delta(|\bm{p}|-\Omega_g)\Big)
    \Big(\int \frac{d^3k}{(2\pi)^3}
    \frac{ \bm{k}\cdot \bm{d} }
    {  \Omega_g^2 - |\bm{k}|^2 } e^{ i \bm{k} \cdot (\bm{y}-\bm{R}) }\Big),
    \label{app:2nd_g}
\end{align}
where, in the last equality, we used 
\begin{align}
    \frac{1}
    { |\bm{p} - \bm{k}|^2 }
     =\int\frac{d^3y}{4\pi |\bm{y}|}  e^{-i(\bm{p}-\bm{k}) \cdot \bm{y}}  .
    \label{app:tec1}
\end{align}
The $\bm{p}$ integral in Eq.\eqref{app:2nd_g} is evaluated as 
\begin{align}
    \int \frac{d^3p}{(2\pi)^3} \ \bm{p} \cdot \bm{d} \
    e^{ - i\bm{p} \cdot (\bm{y} - \bm{R}) } 
    \delta(|\bm{p}| - \Omega_g)
    &=
    \bm{d} \cdot i \nabla_y
    \left(
    \int \frac{d^3p}{(2\pi)^3} \ 
    e^{-i \bm{p}\cdot (\bm{y} - \bm{R})} \
    \delta (|\bm{p}| - \Omega_g)
    \right)
    \notag\\
    &=
    \frac{\Omega_g}{2 \pi^2} \bm{d} \cdot i \nabla_y
    \Big(\frac{\sin(\Omega_g |\bm{y}-\bm{R}|)}{|\bm{y}-\bm{R}|} \Big)
      \notag \\
    &=
    \frac{i [\bm{d} \cdot (\bm{y}-\bm{R})] \Omega_g}{2 \pi^2}  
    \cdot \frac{\Omega_g |\bm{y}-\bm{R}| \cos(\Omega_g |\bm{y}-\bm{R}|) - \sin(\Omega_g |\bm{y}-\bm{R}|)}{|\bm{y}-\bm{R}|^3}. 
    \label{app:p-int}
\end{align}
The same procedure can be also applied for the momentum $\bm{k}$ integral in Eq.\eqref{app:2nd_g}: 
\begin{align}
    \int \frac{d^3k}{(2\pi)^3} 
    \frac
    { 
    \bm{k} \cdot \bm{d} \ 
    }
    { \Omega^2_g - |\bm{k}|^2 }
    \ e^{  i\bm{k} \cdot (\bm{y} - \bm{R}) }
    &=
    -i \bm{d} \cdot
    \nabla_y
    \left(
    \int \frac{d^3k}{(2\pi)^3}
    \frac
    { 
    e^{  i\bm{k} \cdot (\bm{y} - \bm{R}) }
    }
    { \Omega^2_g - |\bm{k}|^2 }
    \right).
    \notag \\ 
    &=
    -i \bm{d} \cdot
    \nabla_y
    \left(\frac{1}{2 \pi^2 |\bm{y}-\bm{R}|}
    \int^\infty_0 dk \ 
    \frac{k\sin(k|\bm{y}-\bm{R}|)}{\Omega^2 - k^2} \right)
    \notag \\
    &=  -i \bm{d} \cdot
    \nabla_y \Big(
    - \frac{\cos(\Omega |\bm{y}-\bm{R}|)}{4 \pi |\bm{y}-\bm{R}|} \Big)
    \notag
    \\
    &=
    - \frac{i[\bm{d} \cdot (\bm{y}-\bm{R})]}{4\pi}
    \frac{\cos(\Omega_g |\bm{y}-\bm{R}|) + \Omega_g |\bm{y}-\bm{R}|  \sin(\Omega_g |\bm{y}-\bm{R}|)}{|\bm{y}-\bm{R}|^3}. 
    \label{app:k-int}
\end{align}
The $k$ integral in the second line is evaluated as a principal value. 
Substituting Eqs.\eqref{app:p-int} and \eqref{app:k-int} into Eq.\eqref{app:2nd_g} and changing the integral variables as $\bm{y} \rightarrow \bm{y}-\bm{R}$, we have the following integral form, 
\begin{align}
    &8 \pi R \Phi (\bm{R}) \Omega_g \int \frac{d^3y}{|\bm{y}|} \Big(\int \frac{d^3p}{(2\pi)^3} \bm{p}\cdot \bm{d} \, e^{-i\bm{p} \cdot (\bm{y}-\bm{R})}  \delta(|\bm{p}|-\Omega_g)\Big)
    \Big(\int \frac{d^3k}{(2\pi)^3}
    \frac{ \bm{k}\cdot \bm{d} }
    {  \Omega_g^2 - |\bm{k}|^2 } e^{ i \bm{k} \cdot (\bm{y}-\bm{R}) }\Big)
    \nonumber 
    \\
    &\quad =\frac{1}{\pi^2} R \Phi (\bm{R}) \Omega^2_g \int \frac{d^3y}{|\bm{y}+\bm{R}|} (\bm{d} \cdot \bm{y})^2  \frac{\Omega_g |\bm{y}| \cos(\Omega_g |\bm{y}|) - \sin(\Omega_g |\bm{y}|)}{|\bm{y}|^3}
    \frac{\cos(\Omega_g |\bm{y}|) + \Omega_g |\bm{y}|  \sin(\Omega_g |\bm{y}|)}{|\bm{y}|^3}
    \nonumber 
    \\
    &\quad 
    =\frac{1}{\pi^2} R \Phi (\bm{R}) \Omega^2_g d_k d_\ell F_{k\ell}(\bm{R}), 
    \label{app:2nd_g2}
\end{align}
where we introduced the function $F_{k\ell}(\bm{R})$ defined by
\begin{align}
    F_{k\ell}(\bm{R})
    \equiv
    \int \frac{d^3y}{|\bm{y} + \bm{R}|}
    \ y_k  y_\ell
    \frac{\Omega_g |\bm{y}| \cos(\Omega_g |\bm{y}|) - \sin(\Omega_g |\bm{y}|)}{|\bm{y}|^3}
    \frac{\cos(\Omega_g |\bm{y}|) + \Omega_g |\bm{y}|  \sin(\Omega_g |\bm{y}|)}{|\bm{y}|^3}. 
    \label{eq:def_tensor-F}
\end{align}
The function $F_{k\ell}$ is covariant under the rotation of the vector $\bm{R}$, so it is expressed by the scalar functions of $R$ as 
\begin{align}
    F_{k\ell}(\bm{R}) =   B_1 (R)\delta_{k\ell} + B_2(R) ( R_k R_\ell - R^2 \delta_{k\ell}).
    \label{eq:re_tensor_F}
\end{align}
Because of Eqs.\eqref{eq:def_tensor-F} and \eqref{eq:re_tensor_F}, it turns out that those scalar function $B_{1,2}(R)$ is given by
\begin{align}
    B_1(R) 
    &=
    \frac{1}{R^2}
    \int \frac{d^3y}{|\bm{y} + \bm{R}|}
    ( \bm{R} \cdot \bm{y} )^2
      \frac{\Omega_g |\bm{y}| \cos(\Omega_g |\bm{y}|) - \sin(\Omega_g |\bm{y}|)}{|\bm{y}|^3}
    \frac{\cos(\Omega_g |\bm{y}|) + \Omega_g |\bm{y}|  \sin(\Omega_g |\bm{y}|)}{|\bm{y}|^3}
    \notag \\
    &=
    \int^{\infty}_0 dy \, y^2 \int^\pi_0 d\theta \, \sin\theta \int^{2\pi}_0 d\varphi \,
    \frac{ y^2 \cos^2\theta  }
    { \sqrt{ y^2 + 2 R \cos\theta + R^2 } }
    \notag \\
    &\quad\times
    \frac{\Omega_g \, y \cos(\Omega_g \, y) - \sin(\Omega_g \, y)}{y^3}
    \frac{\cos(\Omega_g \, y) + \Omega_g \, y  \sin(\Omega_g \, y)}{\, y^3}
    \notag \\
    &=
    - \frac{\pi \Omega_g}{3 R} f_1(R \Omega_g),
    \label{eq:def_B1}
\end{align}
and 
\begin{align}
    B_2(R)
    &=
    \frac{1}{2 R^2}
    \left(
    3 B_1(R) - \delta_{k\ell} F_{k\ell}(\bm{R})
    \right)
    \notag \\
    &=
    \frac{3}{2 R^2} \int \frac{d^3y}{|\bm{y} + \bm{R}|}
    \left[
    \frac{( \bm{R} \cdot \bm{y} )^2}{R^2} - \frac{1}{3} \bm{y}^2 
    \right]
     \frac{\Omega_g |\bm{y}| \cos(\Omega_g |\bm{y}|) - \sin(\Omega_g |\bm{y}|)}{|\bm{y}|^3}
    \frac{\cos(\Omega_g |\bm{y}|) + \Omega_g |\bm{y}|  \sin(\Omega_g |\bm{y}|)}{|\bm{y}|^3}
    \notag \\
    &=
    \frac{3}{2 R^2}
    \int^{\infty}_0 dy \, y^2 \int^\pi_0 d\theta \, \sin\theta \int^{2\pi}_0 d\varphi \,
    \frac{ y^2  }
    { \sqrt{ y^2 + 2 R \cos\theta + R^2 } }
    \left[
    \cos^2\theta - \frac{1}{3} 
    \right]
    \notag \\
    &\quad 
    \times
    \frac{\Omega_g\, y \cos(\Omega_g \, y) - \sin(\Omega_g \, y)}{y^3}
    \frac{\cos(\Omega_g \, y) + \Omega_g \, y  \sin(\Omega_g \, y)}{\, y^3}
    \notag \\
    &=
    - \frac{\pi \Omega_g}{2R^3} f_2 (R \Omega_g),
    \label{eq:def_B2}
\end{align}
where $f_{1,2}(x)$ are defined in Eqs.\eqref{eq:def_f1} and \eqref{eq:def_f2}.
Thus, we have 
\begin{align}
    F_{k\ell}(\bm{R}) =    - \frac{\pi \Omega_g}{3 R} f_1(R \Omega_g)\delta_{k\ell}  - \frac{\pi \Omega_g}{2R^3} f_2 (R \Omega_g) ( R_k R_\ell - R^2 \delta_{k\ell}),
    \label{app:F2}
\end{align}
and then the second term of $\gamma_g$ can be evaluated as
\begin{align}
   &\frac{1}{\pi^2} R \Phi (\bm{R}) \Omega^2_g d_k d_\ell F_{k\ell} 
    \nonumber 
    \\
     &=\frac{1}{\pi^2} R \Phi (\bm{R}) \Omega^2_g d_k d_\ell \Big[ 
    - \frac{\pi \Omega_g}{3 R} f_1(R \Omega_g)\delta_{k\ell}  - \frac{\pi \Omega_g}{2R^3} f_2 (R \Omega_g) ( R_k R_\ell - R^2 \delta_{k\ell})\Big]
    \nonumber 
    \\
     &=-\frac{\Omega^3_g d^2}{6\pi}  \Phi (\bm{R}) \Big[ 
    2 f_1(R \Omega_g)  + 3 \frac{(\bm{d}\cdot \bm{R})^2-d^2 R^2}{d^2R^2}f_2 (R \Omega_g) \Big].
    \label{app:2nd_g3}
\end{align}
Eqs.\eqref{app:1st_g} and \eqref{app:2nd_g3} lead to the following form of $\gamma_g$:  
\begin{align}
    \gamma_g 
    &=[1+4\Phi(\bm{R})]\frac{\Omega^3_g d^2}{6\pi}
     -\frac{\Omega^3_g d^2}{6\pi}  \Phi (\bm{R}) \Big[ 
    2 f_1(R \Omega_g)  + 3 \frac{(\bm{d}\cdot \bm{R})^2-d^2 R^2}{d^2R^2}f_2 (R \Omega_g) \Big]
    \nonumber 
    \\
    &\approx \frac{\Omega^3 d^2}{6\pi} \Big[1+7\Phi(\bm{R})- 2\Phi (\bm{R}) f_1(R \Omega)  - 3 \frac{(\bm{d}\cdot \bm{R})^2-d^2 R^2}{d^2R^2}\Phi(\bm{R}) f_2 (R \Omega) \Big], 
    \label{app:gamma_g2}
\end{align}
where we substituted $\Omega_g=(1+\Phi)\Omega$ into the above and evaluated it up to the first order of $\Phi$ in the last equality. 
This is nothing but Eq.\eqref{eq:gamma_g}.

\section{
Derivation of Eq.\eqref{eq:E_change}
}
\label{app:Derive_E-change}
In this section, we derive Eq.\eqref{eq:E_change}.
To this end, we firstly investigate the dynamics of the scalar field based on the action $S_\text{E}[\phi] +S_\text{int}[\bm{q},\phi]$ given by Eqs.\eqref{eq:def_S_E_curve} and \eqref{eq:def_S_int_curve}. 
Varying the action with respect to $\phi$, we get the following differential equation:
\begin{align}
    \nabla^\mu \nabla_\mu \phi(x)+J(\bm{q};x)=0, 
    \label{app:phi_dyn}
\end{align}
where $J(\bm{q};x)=\sum_{i=1}^{2}\lambda_i \  \frac{d\tau_i}{dt} \, \delta^3 (\bm{x} - \bm{q}_i(t))/\sqrt{-g(x)}$ is the source yielded by the composite system of two particles. 
The solution of the equation is given by the usual Green's function method: 
\begin{align}
    \phi(x)=-\int_{\mathbb{R}^4} d^4y \, \sqrt{-g(y)} G_\text{R}(x,y)J(\bm{q};y), 
    \label{app:phi_sol}
\end{align}
where the retarded Green's function $G_\text{R}$ satisfies 
\begin{align}
      \nabla^\mu \nabla_\mu G_\text{R}(x,y)=\frac{1}{\sqrt{-g(x)}}\delta^4(x-y) 
    \label{app:G_R}
\end{align}
and $G_\text{R}(x-y)=0$ for $x^0-y^0 <0$. 
Substituting the metric $g_{\mu\nu}$ of Eq.\eqref{eq:def_metric} into Eq.\eqref{app:G_R} and expanding it with respect to $\Phi$, we find 
\begin{align}
    \left(
    -\partial^2_0 +  4\Phi(\bm{x}) \partial^2_0+\nabla^2
    \right)
    G_\text{R}(x,y)
    =\delta^4(x-y). 
    \label{app:G_R2}
\end{align}
As mentioned in Appendix \ref{app:Analysys_two-point}, we emphasize the following derivation is valid for a wavelength of scalar field larger than the gravitational radius.
We can perturbatively evaluate the retarded Green's function up to the first order of $\Phi$. 
Under the following form
\begin{align}
    G_\text{R}(x,y) \approx  G_0(x,y) + \delta G(x,y),
    \label{app:G0+delta G}
\end{align}
Eq.\eqref{app:G_R2} can be rewritten as,
\begin{align}
    \left(
    -\partial^2_0 + \nabla^2
    \right)
    G_0(x,y)
    &=
   \delta^4 (x-y),
    \label{app:G0}
    \\
    \left(
    -\partial^2_0 + \nabla^2
    \right)
   \delta G(x,y)
    &=
    -
    4\Phi(\bm{x}) \partial^2_0  G_0(x,y),
    \label{app:deltaG}
\end{align}
where 
$G_0 (x,y)=0$ and $\delta G(x,y)=0$ for $x^0-y^0<0$. 
The $G_0$ is nothing but the retarded Green's function in the flat spacetime given as 
\begin{align}
    G_0 (x,y)=-\frac{\delta(t-\tau-|\bm{x}-\bm{y}|)}{4\pi |\bm{x}-\bm{y}|},
    \label{app:G0_sol}
\end{align}
where $x^\mu=(t,\bm{x})^\text{T}$ and $y^{\mu}=(\tau,\bm{y})^\text{T}$, and hence we find the following $\delta G$, 
\begin{align}
   \delta G(x,y)
    &=
    -
    4\int_{\mathbb{R}^4} d^4 z \, G_0 (x,z)\Phi(\bm{z}) \partial^2_0 G_0(z,y)
    \nonumber
    \\
    &=
    -\frac{1}{4\pi^2}
    \int_{\mathbb{R}^3} d^3 z \frac{\Phi(\bm{z})}{|\bm{x}-\bm{z}||\bm{z}-\bm{y}|} \frac{\partial^2}{\partial t^2}\delta(t-\tau-|\bm{x}-\bm{z}|-|\bm{z}-\bm{y}|)
    . 
    \label{app:deltaG_sol}
\end{align}
From the effective interaction, Eq.\eqref{eq:S_int3}, the source of the scalar field can be approximated as 
\begin{align}
\sqrt{-g(x)} J(\bm{q};x) \approx -[1+\Phi(\bm{R})] \lambda \bm{r} \cdot \nabla_x \delta^3 (\bm{x}-\bm{R}).
    \label{app:J}
\end{align}
Since the COM motion of the composite system is at rest, the scalar field is excited by the internal motion of the system. 
We replace $\lambda \bm{r}(t)$ with the effective dipole $[1+\Phi(\bm{R})] \bm{d}(t)$ to respect the approximation \eqref{eq:lambda_r}. 
The solution $\phi(x)$ is perturbatively given as
\begin{align}
    \phi(x)  
    &\approx -\int_{\mathbb{R}^4} d^4y \,  \Big(-G_0 (x,y)[1+2\Phi(\bm{R})] \bm{d} \cdot \nabla_y \delta^3 (\bm{y}-\bm{R})-\delta G(x,y) \bm{d} \cdot \nabla_y \delta^3 (\bm{y}-\bm{R}) \Big)
    \nonumber 
    \\
    &=-\int^\infty_{-\infty} d\tau \, \Big( [1+2\Phi(\bm{R})]\bm{d}(\tau) \cdot \nabla_R G_0 (t,\bm{x},\tau,\bm{R}) + \bm{d}(\tau) \cdot \nabla_R \delta G(t,\bm{x},\tau,\bm{R})  \Big)
     \nonumber 
    \\
    &=\frac{1}{4\pi} [1+2\Phi(\bm{R})] \Big\{ \frac{\bm{x-R}}{|\bm{x}-\bm{R}|^3} \cdot \bm{d}(t-|\bm{x}-\bm{R}|)+\frac{\bm{x-R}}{|\bm{x}-\bm{R}|^2} \cdot \dot{\bm{d}}(t-|\bm{x}-\bm{R}|) \Big \} 
    \nonumber 
    \\
    &\quad+\frac{1}{4\pi^2} \int_{\mathbb{R}^3} d^3 z \frac{\Phi(\bm{z})}{|\bm{x}-\bm{z}|} 
    \Big \{
    \frac{\bm{z}-\bm{R}}{|\bm{z}-\bm{R}|^3} \cdot  \ddot{\bm{d}}(t-|\bm{x}-\bm{z}|-|\bm{z}-\bm{R}|) + \frac{\bm{z}-\bm{R}}{|\bm{z}-\bm{R}|^2} \cdot  \dddot{\bm{d}}(t-|\bm{x}-\bm{z}|-|\bm{z}-\bm{R}|) \Big\}.
    \label{app:phi_sol3}
\end{align}

Next, we compute the power of the scalar radiation emitted by the internal motion of the composite system. 
The radiation power $P$ is defined by 
\begin{align}
    P= \lim_{r\rightarrow \infty} \int d^2 \hat{r} \, r^{2} \, T^{r0}(\bm{r}),
    \label{eq:P}
\end{align}
where $T^{r0} = -\partial_t \phi(x) \partial_r \phi(x)$ denotes the radial energy flux, and $d^2\hat{r}$ is the solid-angle element associated with the spatial coordinate $\bm{r}$.
By setting $\bm{r} = \bm{x} - \bm{R}$, the solution $\phi$ is written as 
\begin{align}
    \phi(x)  
    &=\frac{1}{4\pi} [1+2\Phi(\bm{R})] \Big\{ \frac{\bm{r}}{r^3} \cdot \bm{d}(t-r)+\frac{\bm{r}}{r^2} \cdot \dot{\bm{d}}(t-r) \Big \} 
    \nonumber 
    \\
    &\quad+\frac{1}{4\pi^2} \int_{\mathbb{R}^3} d^3 z \frac{\Phi(\bm{z}+\bm{R})}{|\bm{r}-\bm{z}|} 
    \Big \{
    \frac{\bm{z}}{z^3} \cdot  \ddot{\bm{d}}(t-|\bm{r}-\bm{z}|-z) + \frac{\bm{z}}{z^2} \cdot  \dddot{\bm{d}}(t-|\bm{r}-\bm{z}|-z) \Big\},
    \label{app:phi_sol4}
\end{align}
where we changed the integral variables as $\bm{z} \rightarrow \bm{z}-\bm{R}$, and $z=|\bm{z}|$. 
Since we are interested in the scalar radiation dissipating sufficiently far away, we should consider the limit that $r=|\bm{r}|$ is very large. 
In the large $r$, the solution of $\phi$ is
\begin{align}
    \phi(x)  
    &\approx \frac{1}{4\pi r} [1+2\Phi(\bm{R})] \, \hat{r} \cdot \dot{\bm{d}}(t-r) 
    \nonumber 
    \\
    &\quad+\frac{1}{4\pi^2r} \int_{\mathbb{R}^3} d^3 z \Phi(\bm{z}+\bm{R}) 
    \Big \{
    \frac{\bm{z}}{z^3} \cdot  \ddot{\bm{d}}(t-r-z+\hat{r} \cdot \bm{z}) + \frac{\bm{z}}{z^2} \cdot  \dddot{\bm{d}}(t-r-z+\hat{r} \cdot \bm{z}) \Big\},
    \label{app:phi_sol5}
\end{align}
where $\hat{r}=\bm{r}/r$. 
Because of this result,  the radial energy flux in the large $r$ is evaluated up to the first order of $\Phi$ as 
\begin{align}
    T^{r0} 
    &= -\partial_t \phi(x) \partial_r \phi(x)
    \notag
    \\
    &\approx
   \Big[\frac{1}{4\pi r} [1+2\Phi(\bm{R})] \, \hat{r} \cdot \ddot{\bm{d}}(t-r) 
    \nonumber 
    \\
    &\quad+\frac{1}{4\pi^2r} \int_{\mathbb{R}^3} d^3 z \Phi(\bm{z}+\bm{R}) 
    \Big \{
    \frac{\bm{z}}{z^3} \cdot  \dddot{\bm{d}}(t-r-z+\hat{r} \cdot \bm{z}) + \frac{\bm{z}}{z^2} \cdot  \ddddot{\bm{d}}(t-r-z+\hat{r} \cdot \bm{z}) \Big\} \Big]^2
    \nonumber 
    \\
    &
    \approx \frac{1}{16\pi^2 r^2} [1+4\Phi(\bm{R})] \, [\hat{r} \cdot \ddot{\bm{d}}(t-r)]^2
    \nonumber 
    \\
    &\quad+\frac{1}{8\pi^3r^2} \hat{r} \cdot \ddot{\bm{d}}(t-r)\int_{\mathbb{R}^3} d^3 z \Phi(\bm{z}+\bm{R}) 
    \Big \{
    \frac{\bm{z}}{z^3} \cdot  \dddot{\bm{d}}(t-r-z+\hat{r} \cdot \bm{z}) + \frac{\bm{z}}{z^2} \cdot  \ddddot{\bm{d}}(t-r-z+\hat{r} \cdot \bm{z}) \Big\}. 
    \label{app:Tr0}
\end{align}
Here, we assume the oscillating dipole 
\begin{align}
  \bm{d}(t)=\bm{d} \cos(\Omega_g t),
\end{align}
where $\Omega_g =[1+\Phi(\bm{R})]\Omega$. 
To evaluate the power $P$, we perform a time average and neglect the oscillatory terms with the period $2\pi/\Omega_g$ from the radial energy flux. 
Practically, ignoring $\cos(2\Omega_g (t-r))$,  $\sin[ 2\Omega_g (t-r) + \Omega_g(\hat{\bm{r}} \cdot \bm{x'} - |\bm{x'}| )]$ and $\cos[ 2\Omega_g (t-r) +\Omega_g( \hat{\bm{r}} \cdot \bm{x'} - |\bm{x'}| )]$ appearing in \eqref{app:Tr0}, we get the following radial energy flux, 
\begin{align}
    T^{r0} 
    &=\frac{\Omega^4_g}{32\pi^2 r^2} [1+4\Phi(\bm{R})] \, (\hat{r} \cdot \bm{d})^2
    \nonumber 
    \\
    &\quad-\frac{1}{16\pi^3r^2} \hat{r} \cdot \bm{d}\int_{\mathbb{R}^3} d^3 z \Phi(\bm{z}+\bm{R}) 
    \Big \{
    \frac{\bm{z} \cdot \bm{d}}{z^3} \Omega^5_g \sin[\Omega_g (\hat{r}\cdot \bm{z}-z)] + \frac{\bm{z} \cdot \bm{d}}{z^2} \Omega^6_g  \cos[\Omega_g (\hat{r}\cdot \bm{z}-z)]\Big\}. 
\end{align}
Using the following relations,
\begin{align}
    & \int d^2\hat{r} \, ( \hat{r} \cdot \bm{d} )^2 = \frac{4 \pi d^2}{3},
    \notag
    \\
    & \int d^2 \hat{r} \, \hat{r} \cdot \bm{d} \, \sin[\Omega_g (\hat{r}\cdot \bm{z}-z)]
    = -4\pi \frac{\bm{z} \cdot \bm{d}}{z}  \Big[\cos(\Omega_g z)-\frac{\sin(\Omega_g z)}{\Omega_g z} \Big] \frac{\cos(\Omega_g z)}{\Omega_g z},
    \notag 
    \\
    &  \int d^2 \hat{r} \,\hat{r} \cdot \bm{d} \, \cos[\Omega_g (\hat{r}\cdot \bm{z}-z)]
    = -4\pi \frac{\bm{z} \cdot \bm{d}}{z} \Big[\cos(\Omega_g z)-\frac{\sin(\Omega_g z)}{\Omega_g z} \Big] \frac{\sin(\Omega_g z)}{\Omega_g z},
    \notag
\end{align}
we can get the radiation power $P$ as 
\begin{align}
   P&= \lim_{r\rightarrow \infty} \int d^2 \hat{r} \, r^{2} \, T^{r0} 
   \nonumber 
   \\
   &=\frac{\Omega^4_g}{32\pi^2 } [1+4\Phi(\bm{R})] \, \frac{4 \pi d^2}{3}
    \nonumber 
    \\
    &\quad-\frac{1}{16\pi^3} \int_{\mathbb{R}^3} d^3 z \Phi(\bm{z}+\bm{R}) 
    \Big \{
    \frac{\bm{z} \cdot \bm{d}}{z^3} \Omega^5_g \Big(-4\pi \frac{\bm{z} \cdot \bm{d}}{z}  \Big[\cos(\Omega_g z)-\frac{\sin(\Omega_g z)}{\Omega_g z} \Big] \frac{\cos(\Omega_g z)}{\Omega_g z} \Big)
    \nonumber 
    \\
    &
    \quad 
    + \frac{\bm{z} \cdot \bm{d}}{z^2} \Omega^6_g  \Big(-4\pi \frac{\bm{z} \cdot \bm{d}}{z}  \Big[\cos(\Omega_g z)-\frac{\sin(\Omega_g z)}{\Omega_g z} \Big] \frac{\sin(\Omega_g z)}{\Omega_g z} \Big)\Big\}
    \nonumber 
    \\
    &=\frac{\Omega^4_g d^2}{24\pi } [1+4\Phi(\bm{R})] 
    +\frac{\Omega^3_g}{4\pi^2} \int_{\mathbb{R}^3} d^3 z \Phi(\bm{z}+\bm{R})(\bm{z} \cdot \bm{d})^2 
    \frac{\Omega_g z\cos(\Omega_g z)-\sin(\Omega_g z)}{z^3}  \frac{
    \cos(\Omega_g z) 
    +  \Omega_g z  \sin(\Omega_g z) }{z^3}. 
    \label{app:P2}
\end{align}
Substituting $\Phi(\bm{z}+\bm{R})=-GM/|\bm{z}+\bm{R}|=R \Phi(\bm{R})/|\bm{z}+\bm{R}|$ into the second term of $P$, we can perform the $\bm{z}$ integral as
\begin{align}
   (\text{Second term of } P)&=\frac{\Omega^3_g}{4\pi^2} \int_{\mathbb{R}^3} d^3 z \Phi(\bm{z}+\bm{R})(\bm{z} \cdot \bm{d})^2 
    \frac{\Omega_g z\cos(\Omega_g z)-\sin(\Omega_g z)}{z^3}  \frac{
    \cos(\Omega_g z) 
    +  \Omega_g z  \sin(\Omega_g z) }{z^3}
    \nonumber 
    \\
    &=\frac{\Omega^3_g R \Phi(\bm{R})}{4\pi^2} \int_{\mathbb{R}^3} \frac{d^3 z}{|\bm{z}+\bm{R}|} (\bm{z} \cdot \bm{d})^2 
    \frac{\Omega_g z\cos(\Omega_g z)-\sin(\Omega_g z)}{z^3}  \frac{
    \cos(\Omega_g z) 
    +  \Omega_g z  \sin(\Omega_g z) }{z^3}
     \nonumber 
    \\
    &=\frac{\Omega^3_g R \Phi(\bm{R})}{4\pi^2} d_k d_\ell F_{k\ell}(\bm{R}) 
    \nonumber 
    \\
    &= \frac{\Omega^3_g R \Phi(\bm{R})}{4\pi^2} d_k d_\ell \Big[- \frac{\pi \Omega_g}{3 R} f_1(R \Omega_g)\delta_{k\ell}  - \frac{\pi \Omega_g}{2R^3} f_2 (R \Omega_g) ( R_k R_\ell - R^2 \delta_{k\ell}) \Big]
    \nonumber 
    \\
    &=- \frac{\Omega^4_g d^2 \Phi(\bm{R})}{24\pi} \Big[2 f_1(R \Omega_g)  + 3 \frac{ (\bm{d} \cdot \bm{R})^2 - d^2 R^2 }{d^2 R^2} f_2 (R \Omega_g)\Big], 
    \label{eq:P_2nd}
\end{align}
where we used $F_{k\ell}$ defined in \eqref{eq:def_tensor-F} in the third equality  and substituted its explicit form \eqref{app:F2} in the fourth equality, respectively. 
Using the form of the second term of $P$, we finally find the following $P/\Omega_g$ up to $O(\Phi)$ as 
\begin{align}
   \frac{P}{\Omega_g}&=\frac{\Omega^3_g d^2}{24\pi } [ 1+4\Phi(\bm{R})] 
    - \frac{\Omega^3_g d^2 \Phi(\bm{R})}{24\pi} \Big[2 f_1(R \Omega_g)  + 3 \frac{ (\bm{d} \cdot \bm{R})^2 - d^2 R^2 }{d^2 R^2} f_2 (R \Omega_g)\Big] 
    \nonumber 
    \\
    & \approx 
    \frac{\Omega^3 d^2}{24\pi }\Big[ 1+7\Phi(\bm{R}) -2 \Phi(\bm{R}) f_1(R \Omega)   - 3 \frac{(\bm{d} \cdot \bm{R})^2 - d^2 R^2 }{d^2 R^2} \Phi(\bm{R}) f_2 (R \Omega)\Big]
    \nonumber 
    \\
    & = \frac{1}{4} \gamma_g.
    \label{app:P3}
\end{align}
Hence, the energy dissipation rate of the dipole, $P/\Omega_g$, is proportional to the spontaneous emission rate $\gamma_g$ given in \eqref{eq:gamma_g}.  

\end{appendix}

\end{document}